\documentclass[12pt]{article}

\pdfoutput=1 
\usepackage[applemac]{inputenc}
\usepackage[T1]{fontenc}
\usepackage{amsmath}
\usepackage{amsfonts}

\usepackage[english]{babel}

\usepackage{geometry}                % See geometry.pdf to learn the layout options. There are lots.
\usepackage[parfill]{parskip}    % Activate to begin paragraphs with an empty line ra than an indent
\usepackage{graphicx}
\usepackage{amssymb}
\usepackage{epstopdf}
\usepackage[all]{xy}

\usepackage{mathrsfs}
\usepackage{hyperref}
\usepackage{enumitem}

\usepackage{amsthm} 
\theoremstyle{definition}

\theoremstyle{remark} 
 
\theoremstyle{plain}

\theoremstyle{plain}

\begin{document}

\title{Cosmology and philosophy}

\author{Daniel Parrochia}
\date{Universit\'{e} of Lyon (France)}
\maketitle

\begin{figure}[h] %  figure placement: here, top, bottom, or page
    \vspace{-1\baselineskip}
	   \hspace{8.3\baselineskip}	  
	   \includegraphics[width=2.5in]{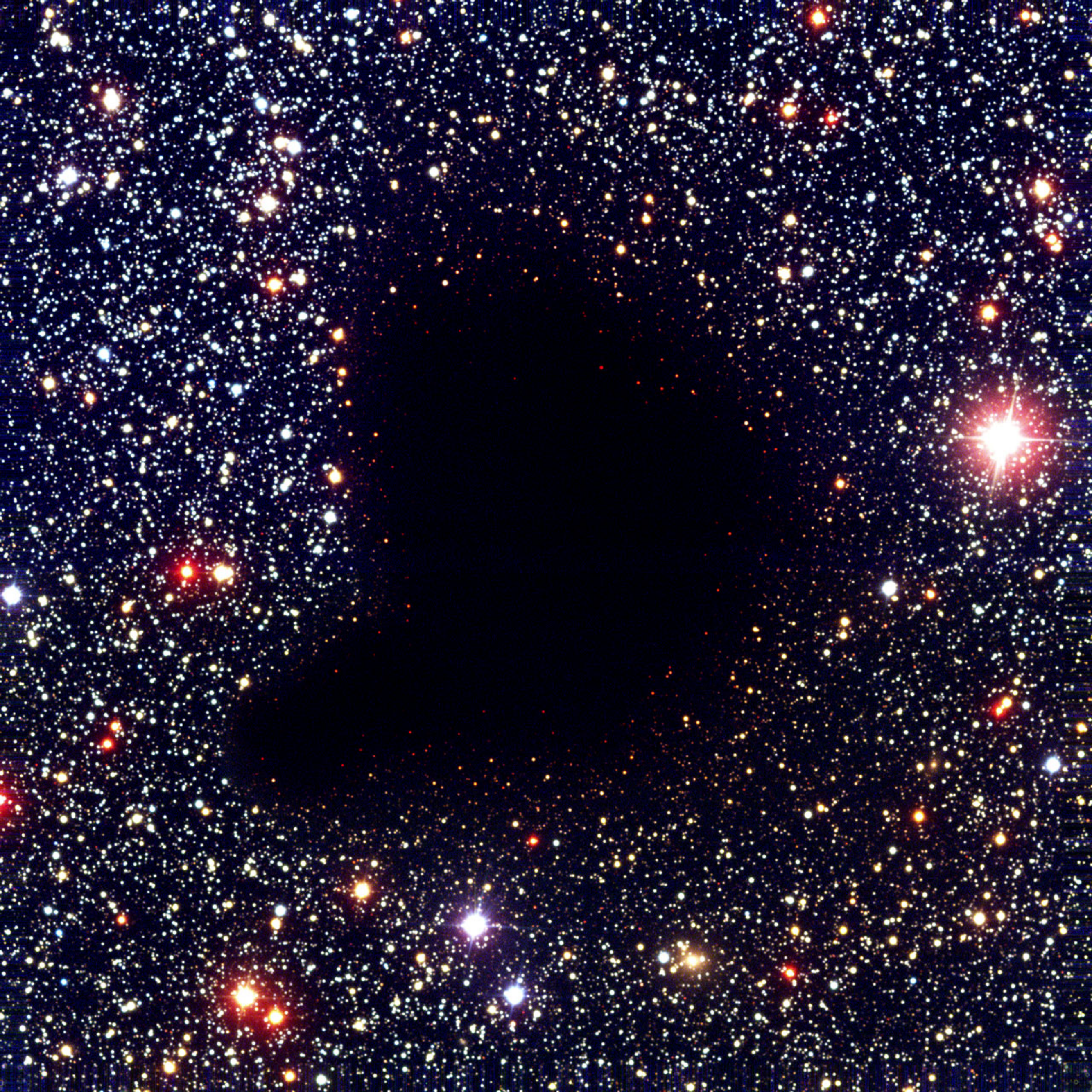} 
	      \vspace{0\baselineskip}
	   \caption{Black cloud Barnard 68 (ESO)}
	   \label{fig: conj11}
	\end{figure}

\textbf{Abstract.}

Scientific cosmology has now reached its period of maturity with the establishment of a standard model, which is the theory of an expanding universe. The question of whether this expansion resolves itself, in the past, into a singularity identifiable with an absolute beginning, or whether the universe in which we are is only one of the multiple possible universes existing either in space or in time, is still under debate. Moreover, the assimilation of the beginning of the universe to a "creation" has often been contested by theology, which, since Thomas Aquinas, if not since the Fathers of the Church, tends to carefully distinguish the two. In the following article, after briefly summarizing some points in the recent history of scientific cosmology, we will attempt to present in broad outline the standard model that scientists have arrived at. Then, we will undertake to study some of the problems it raises as well as the alternative theories that can be opposed to it. Finally, we will discuss the problematic links that scientific cosmology continues to maintain with philosophy and theology, notably the thorny question of creation from nothing ({\it creatio ex nihilo}).


\section{The Contribution of Scientific Cosmology}

While classical Newtonian physics, derived from Galileo, did not, strictly speaking, have a cosmology -- in the sense that it did not imply, in its equations, any particular consequences for the shape of the world or its mathematical-mechanical representation (other than an infinite and eternal space) -- the physics derived from the theory of general relativity\footnote{General relativity, as a theory of gravitation, remained, until recent years, one of the very solid foundations of cosmology. While, on small scales, it has long been questioned -- due to its fundamental incompatibility with quantum mechanics -- it was nevertheless considered, until recently, a good description of the Universe on a large scale. Only recently has a 1\% attenuation of gravity been demonstrated at the cosmological "super-horizon," that is, at scales much larger than the current Hubble radius. If this were true, it would require a slight modification of Einstein's theory for these scales, probably involving cold dark matter $\Lambda$; this "codicil" does not, for the moment, lead to an overhaul (see \cite{Wen}).} -- Einstein's theory led, quite naturally, to cosmologies that were often finitist, thus reconnecting with pre-Socratic speculations or mythological concerns in general.

However, unlike the latter, it developed on the basis of essentially mathematical and physical considerations, and according to the usual scientific methodology, namely that any theoretical construction must be validated or refuted by observations. The notion of "universe" has, as a result, become much more refined. In cosmology, we speak of a very particular universe, a universe that is directly or indirectly "observable," and therefore subject to scientific, theoretical, and experimental investigation.

In the first half of the 20th century, however, the possibility of thinking about the universe as a whole was still contested by certain epistemologists such as G. Bachelard (see \cite{Bac} and our comments in \cite{Bar2}, 235-247), probably due to excessive uncertainty about the possible models of the universe, which were quite numerous at the time, and at a time when no consensus had yet emerged in favor of the expansion theory. But progress in the adjustment of theories and data has been such that it is no longer possible today to dispute the relevance of this bold transgression, which consisted of scientifically considering the universe as a whole.

\subsection{From Pioneers of Expansion to Inflation Theory}

Scientific cosmology developed in the 20th century through a series of theories, which Jacques Merleau-Ponty (see \cite{Mer}) surveyed up until the 1960s. However, it owed its emergence, as he wrote, essentially to the conjunction of a genius physicist (Einstein) and a gigantic telescope (the 2.5-meter aperture telescope on Mount Wilson), which enabled Hubble, an astronomer "in his own right," to derive his famous law of recession of galaxies $V = H_{0}D$, which showed that their speed of receding was proportional to their distance. The law, difficult to establish in 1929, is currently well verified (see Fig. 2). The only remaining problem -- but it may have significant consequences in the long term -- is the question of evaluating the Hubble constant $H_0$. Estimates based on standard candle methods (Cepheid stars, etc.) are concentrated around 73 or 74 km/s/Mpc. In contrast, values derived from observations of the cosmic microwave background using the Lambda CDM model seem to agree more closely around 67 km/s/Mpc. The difference exceeds the margins of error by about 6 sigma. \begin{figure}[h] % figure placement: here, top, bottom, or page
\vspace{-0.5\baselineskip}
\hspace{3\baselineskip}
\includegraphics[width=5in]{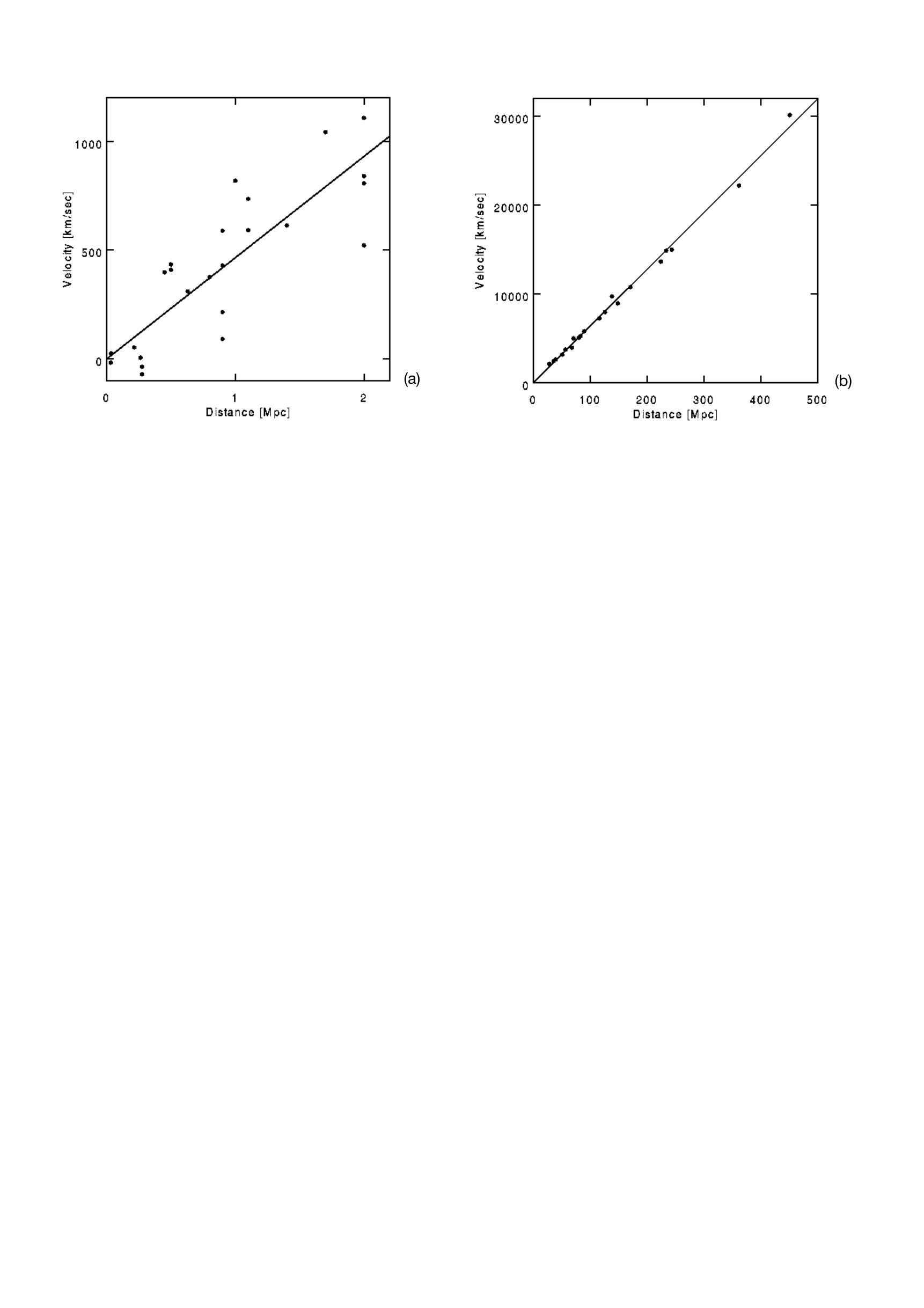}
\vspace{-24\baselineskip}
\caption{Hubble's law in 1929 (a) and in 1996 (b)}
\label{fig: conj111}
\end{figure}

The study of the universe as a whole was therefore able to develop on this basis, and initially as a typically relativistic cosmology, in the wake of Einstein and De Sitter. Friedman and Lema\^{i}tre (see the collection of their main articles introduced by J.-P. Luminet (\cite{Fri}) were the first to find non-static solutions to the equations of General Relativity, while Robertson interpreted the Doppler effect in terms of distance. Then the English school made major contributions with Gamow, Eddington, Hoyle, sometimes including philosophical considerations (as with Tolman, Milne or Whitrow). Specific variants were then explored (theory of large numbers of Eddington, Dirac or Jordan, rotating universe of G\"{o}del, non-homogeneous isotropic models or lattice universe of Lindquist and Wheeler). The great opposition of the time was between the theories of an expanding universe and the so-called stationary state theories defended by Sciama and Hoyle, more skeptical about the interpretation of the redshift of the spectra of distant galaxies, and less inclined to admit without discussion Hubble's law, the data still remaining quite limited.

Gamow and his colleagues -- even though they did not initially believe in the "Big Bang" theory -- were soon to confirm it. By interpreting the formation of the Universe in the Hebrew fashion, that is, not as a "creatio ex nihilo" but as a creation from a prior matter, they showed that the assumption of an initial mixture called {\it Ylem} made it possible to account for the formation of the nuclei of the lightest elements (deuterium, helium and lithium) during the first three minutes of the universe, at a time when the temperature was around 10 billion degrees. They also predicted that once the Universe had cooled, it must have become transparent, and Alpher and Hermann calculated that an echo of the Big Bang in the form of blackbody radiation should now be found at a temperature of around 5 K. This background radiation, which Penzias and Wilson discovered by chance in 1956 while listening to the first Russian Sputnik using a large antenna, was interpreted by Peebles, Roll, and Wilkinson in 1965 as being precisely the residue predicted by Gamow and his colleagues.

From both observational and theoretical physical data, a completely precise model of the universe gradually emerged, whose predictions now correspond perfectly with experimental data, particularly satellite data. Lema\^{i}tre's theory of the primitive atom -- mocked by Hoyle and derisively dubbed the "big bang" theory -- became more widely accepted, particularly after Fran\c{c}ois Englert and Alan Guth independently introduced the idea of an inflationary phase of the Universe to adjust the expansion to the given data. Andrei Linde would later become a proponent of this model, in connection with the development of a multiverse quantum cosmology (see \cite{Lin}). A "quantum foam" then replaced Gamow's {\it Ylem}.

\subsection{The Standard Model}

What can reasonably be called the "standard model of cosmology" is the model that currently most satisfactorily describes the major stages in the history of the observable Universe, as well as its current contents, as revealed by astronomical observations. According to this model, the Universe we inhabit appears as a homogeneous and isotropic expanding space, in which large structures are superimposed, formed by the gravitational collapse of primordial inhomogeneities, themselves formed during an initial phase of expansion called "inflation" because of its rapidity. This standard model of cosmology emerged shortly before the year 2000, following the arrival of a significant amount of astronomical observations, including new galaxy catalogs such as SDSS and 2dFGRS, increasingly detailed observations of anisotropies in the cosmic microwave background with the BOOMERanG and Archeops experiments, and later data from the Wilkinson Microwave Anisotropy Probe (WMAP) space observatory, as well as observations of distant supernovae and gravitational shear effects.

\subsubsection{Data to be satisfied}

This standard model, denoted $\Lambda$CDM (initials of Lambda Cold Dark Matter), also called the "concordance model," is a "Big Bang"-type model parameterized by a cosmological constant denoted by the Greek letter $\Lambda$ and associated with the existence of "dark energy." It is the simplest model that accounts for the properties of the observable Universe, as clarified by WMAP, the COBE and then PLANCK satellites, namely:

1. The existence and structure of the cosmic microwave background;

2. The large-scale structure of the distribution of galaxies (Hubble's law);

3. The abundance of nucleons and light elements (hydrogen, helium, and lithium).

4. The expansion of the Universe and its acceleration.

\subsubsection{Fundamental Assumptions}

Emerged in the late 1990s, after a period when several observed properties of the Universe seemed mutually incompatible, and when no consensus existed on the composition of the Universe's energy densities, this standard $\Lambda$CDM model is based on three assumptions:

($H_{1}$) The cosmological principle, according to which the Universe is homogeneous and isotropic on large scales;

($H_{2}$) The universality principle, according to which gravitation is described by general relativity at all scales;

($H_{3}$) The matter content of the Universe, given by cold dark matter (CDM), baryons, and radiation.

The Universe, moreover, is believed to contain dark energy, the Greek letter $\Lambda$ -- usually the symbol for the cosmological constant -- being the simplest expression.

\subsubsection{Cosmological Parameters}

Replacing the SCDM (Standard Cold Dark Matter) model, developed in the 1990s, the current $\Lambda$CDM model arises from the combination of several observations that constrain certain cosmological parameters:

a) the indirect detection of dark matter, through its gravitational influence within galaxies and galaxy clusters;

b) the estimation of the density of this dark matter, which is less than the critical density of the Universe;

c) constraints on the spatial curvature of the Universe, which indicate that its total density is very close to the critical density;

d) observing the acceleration of the Universe's expansion by studying the luminosity distance of Type Ia supernovae, which implies the existence of dark energy.

The combination of these constraints makes it necessary to add to dark matter a certain form of energy -- "dark" energy -- which has a repulsive effect on the Universe's expansion.

The minimal CDM model -- known as the "vanilla" model -- is more precisely defined by six parameters with independent effects, namely:

1. The baryonic density ($\Omega_{b}$);

2. The dark matter density ($\Omega_{c}$);

3. The dark energy density ($\Omega_{\Lambda}$));

4. The spectral index of the scalar primordial perturbations ($n_{s}$);

5. The amplitude of the curvature primordial perturbations ($\Delta_{R}^2$);

6. The optical reionization thickness ($\tau$).

In this model, the photon density is fixed by the measured temperature of the cosmic microwave background radiation, neutrinos are assumed to have zero mass, and the current Universe as a whole is assumed to be "flat." The remarkable thing is that this model "fits" the observational data quite well (see Fig. 3):

\begin{figure}[h] % figure placement: here, top, bottom, or page
\vspace{-1\baselineskip}
\hspace{5\baselineskip}
\includegraphics[width=8in]{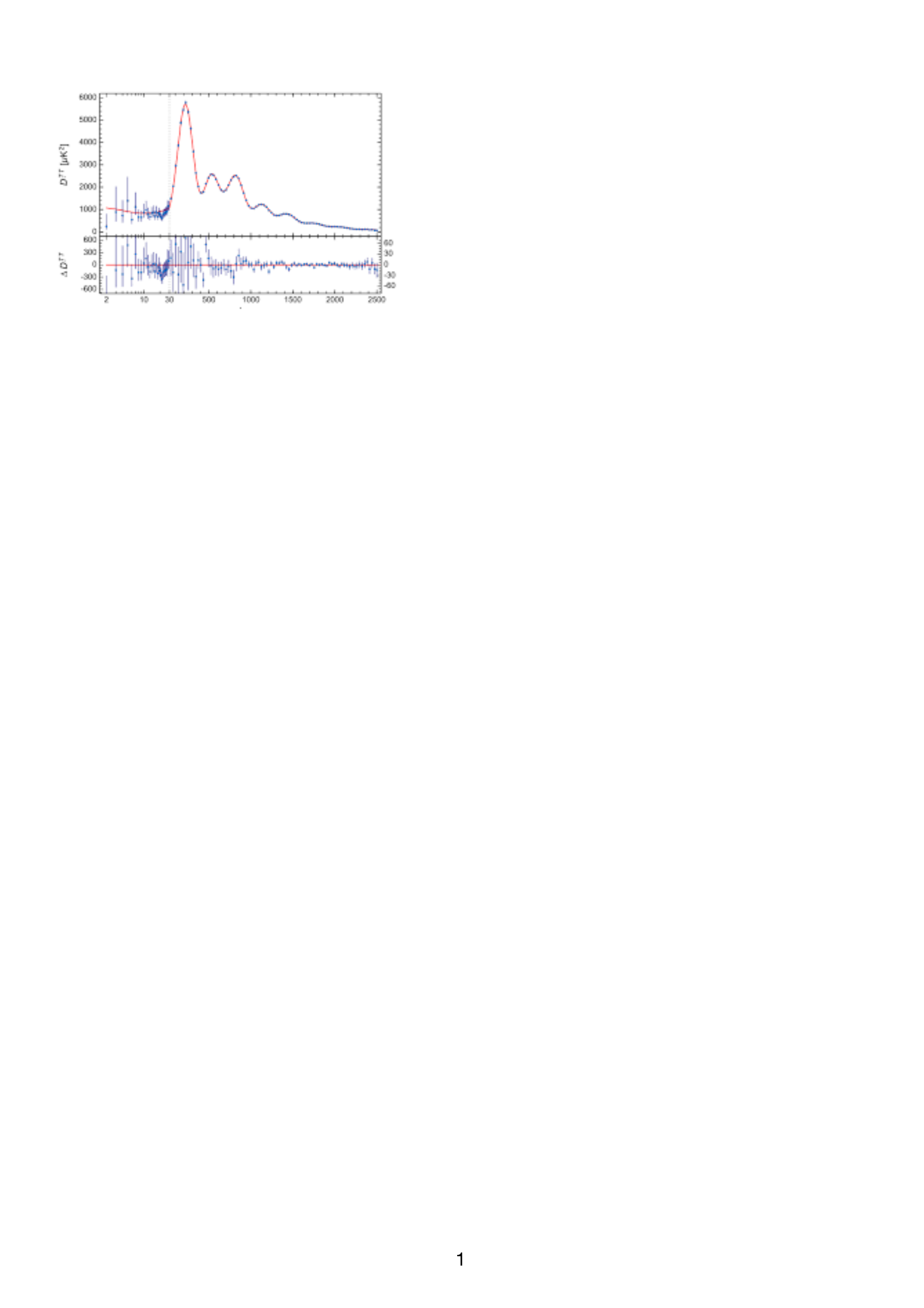}
\vspace{-44\baselineskip}
\caption{Power spectrum of the diffuse background radiation (Planck data) compared to the model}
\label{fig: conj111}
\end{figure}

\subsubsection{A Critical Realism}

This standard model, although favored by the majority of physicists, is nevertheless criticized for its ad hoc assumptions concerning known cosmological problems that are not explained satisfactorily: the problem of structure formation, the problem of flatness, and the asymmetry baryonic equations, the missing baryon problem, the problem of galaxy rotation, the problem of the accelerating expansion of the Universe...

Various variants of this model therefore exist, very often inspired by general relativity (MOND theories, Milne Universe, bi-metric cosmological models, string theories, etc.), without any of them gaining the support of the majority of physicists for the time being.

Things are therefore not definitively settled regarding cosmology, and the image it gives us of the universe may still change.

Nevertheless, the $\Lambda$CDM model can be considered, given current knowledge, a "realistic" model corresponding to an expanding universe. The dynamics of this expansion are governed by general relativity, or possibly another relativistic theory of gravitation. But, in any case, it predicts that if the universe is, as observed, homogeneous and isotropic, then it generally possesses dynamics (it is either expanding or contracting). The observed expansion of the present universe indicates that it was denser and hotter in the past. Several indications, notably the abundance of light elements (from primordial nucleosynthesis), suggest that this hot phase reached a temperature of at least 1 billion degrees Celsius: this is the famous "Big Bang."

The present universe contains a large number of structures such as stars, galaxies, galaxy clusters and superclusters: on a large scale, it is generally homogeneous, but rather irregular on a small scale. Observation of the universe 380,000 years after the "Big Bang", via the cosmic microwave background, shows, however, that the universe at that time was much more homogeneous than it is today. The mechanism of gravitational instability then makes it possible to explain how astrophysical objects could possibly form from an initially relatively inhomogeneous distribution of matter, the attractive effect of gravity tending to help regions denser than their environment to attract neighboring matter. It was therefore gradually that the large structures of the universe formed. The details of this formation process, however, depend on many parameters, in particular the properties of the different forms of matter that fill the universe.

\subsubsection{The Preferred Scenario}

The objective of a realistic cosmological model is to propose a scenario that accounts as accurately as possible for all the observations, which, in this case, highlight two stages:

1) A first stage, which falls under what is now called "primordial cosmology" (see \cite{Pet}), and which must explain:

-- how during the Big Bang the universe could have been in the very homogeneous state observed by the cosmic microwave background radiation;

-- why small irregularities already existed at that time;

-- how the different forms of matter we know (baryonic matter (i.e., atoms), neutrinos, photons) could have emerged from the Big Bang.

2) A second step, which falls under observational cosmology, must explain:

-- the current distribution of galaxies, clusters, and superclusters, revealed by galaxy catalogs;

-- their physical properties (size, mass, temperature, etc.);

-- the evolution of their distribution, which can be observed by comparing the current distribution of these objects with their past distribution, based on observations of earlier periods in the history of the universe.

\subsubsection{Observational Tests}

Today, a number of observations allow us to test the chosen cosmological model:

1. The first is the measurement of the abundance of light elements: it is established that when the temperature of the Universe drops below a billion degrees, the protons and neutrons that exist at that time combine to sometimes form a few nuclei of low-number atoms: deuterium, helium, and lithium: this is primordial nucleosynthesis. The relative abundance of these so-called "light" elements depends on the physical conditions prevailing at that time;

2. The second important element is the existence of the cosmic microwave background, which offers a kind of snapshot of the universe at the time when it became sufficiently sparse for light to propagate freely (the so-called recombination epoch, approximately 380,000 years after the Big Bang); 3. Observations of galaxies, galaxy clusters, and superclusters located at different distances and therefore seen at different times form a third body of evidence supporting the model. It is also supported by the measurement of their gravitational influence on their environment through gravitational lensing effects,
the measurement of the absorption of radiation from distant quasars by non-concentrated matter in galaxies (Lyman-alpha forest),
and the measurement of the expansion rate of the universe (the Hubble constant), as well as its evolution over time through the luminosity distance of certain objects such as Type Ia supernovae.

4. In the more or less distant future, it is possible to observe other phenomena that would allow us to probe other aspects of the observable universe:

-- A first example is the cosmic neutrino background radiation, which is the equivalent of the cosmic microwave background radiation, but for neutrinos. This was emitted much earlier in the history of the universe, a fraction of a second after the Big Bang:

-- A second test would be the cosmic gravitational wave background radiation, which represents primordial gravitational waves from the time when quantum gravity effects occurred in the universe. It would allow us to probe the universe at even earlier times.

\subsubsection{The Enigma of the Dark Zones}

Despite criticism, the standard model -- once again, taking into account current knowledge -- seems to be the best fit for what exists. Obviously, it also leaves us with a number of puzzles, notably the nature and distribution of the different elements it contains. Currently, we have the following distribution:

Visible ordinary matter: 4.9%.

Dark matter: 26.8%

Dark energy: 68.3%.

In summary, the $\Lambda$CDM model represents a universe:

-- spatially homogeneous and isotropic on a large scale (therefore also with constant spatial curvature) corresponding to the Friedmann-Lema\^{i}tre-Robertson-Walker metric,
filled with a perfect fluid with generally zero pressure $p$ -- galactic gas corresponding to: $w = \frac{pc^{-2}}{\rho} = 0$ and with density $\rho$ composed of hot matter (relativistic or radiation) and cold matter (non-relativistic);

-- with zero spatial curvature, in other words, with a curvature parameter $k = 0$ (a sign of a flat universe);

-- which would contain, in addition to ordinary matter, dark matter (excess gravity that galaxies need to keep themselves from breaking apart during their rotation) and dark energy (a global repulsive force that tends to accelerate the expansion of the universe and requires a cosmological constant $\Lambda > 0$), matter and energy about which we know virtually nothing;

-- resulting from a primordial explosion (Big Bang model requiring: $w = - 1/3$).

\section{Problems of the Standard Model and Pre-Big Bang Scenarios}

The Standard Model is verified in terms of basic data on a number of significant points. On other levels, however, it faces several difficulties.

\subsection{Early Galaxies and Supermassive Black Holes}

\subsubsection{The Question of Early Galaxies}

Galaxies with stellar masses reaching about $10^{11}$ solar masses have recently been identified with Doppler redshifts $z$ around 6, about 1 billion years after the Big Bang. These observations were made using the James Webb Space Telescope (JWST) to search for intrinsically red galaxies during the first 750 million years of cosmic history. In this study area, six candidate massive galaxies (stellar mass greater than $10^{10}$ solar masses) could be found at $7.4 \leq z \leq 9.1$, 500-700 billion years after the Big Bang, including one galaxy with a possible stellar mass of about $10^{11}$ solar masses. If verified by spectroscopy, the stellar mass density in these massive galaxies would be much higher than expected (see \cite{Lab}).

\subsubsection{Supermassive Black Holes in the Early Universe}

Since the discovery of "occluded stars" by Schwartzschild in 1915 and their naming as "black holes" by Wheeler in 1967, our understanding of these objects has progressed considerably. In particular, the similarity in the behavior of the hole's area and entropy, both quantities tending to increase irreversibly (see \cite{Bek}), has led to the consideration of the area of the event horizon $A$ as an appropriate measure of the entropy content $S$ of the black hole, giving rise to the very simple Bekenstein-Hawking formula, where $S \approx (1/4) A$.
The entropy of a black hole increases when it absorbs an object, and this increase in entropy is always greater than that of the absorbed object. Similarly, the entropy of a black hole tends to decrease when it "radiates," because this radiation -- discovered by Hawking (see \cite{Haw8}) -- is accompanied by a loss of energy and, consequently, a decrease in the size of the hole. Since the entropy of the radiation emitted by the black hole is seven times greater than the entropy lost, the total entropy (the sum of ordinary entropy and that of the black hole) always increases over time.
For all these reasons, it follows that "young" black holes, existing at the beginning of the universe, should not be enormous.

However, in recent years, larger-than-expected black holes have been discovered in the early universe. For example, in 2017, a black hole weighing 800 million solar masses was discovered in a galaxy located 13.1 billion light-years from Earth. This giant, named J1342+0928, was at the time the most distant black hole ever observed, and therefore the oldest known, since the light from the quasar orbiting it took a little over 13 billion years to reach Earth. This black hole was observed as it was only 690 million years after the Big Bang, that is to say, at a time when the Universe was still in its infancy. Such an observation, likely to overturn current theories on the formation of the Universe, that of black holes and galaxies (\cite{Ban}), is not isolated. Generally speaking, the existence of black holes with a mass more than a billion times greater than that of the Sun, but formed less than a billion years after the Big Bang, has been highlighted several times. It remains difficult to explain. It is possible that these supermassive black holes were born from the collapse of enormous gas clouds with a mass of 10,000 to 1 million times that of the Sun, which existed before the appearance of galaxies. Another hypothesis proposes that these black holes were the product of supernovae of giant stars, with a mass up to about 100 times that of the Sun, and that they then expanded at a surprisingly rapid rate. These primordial black holes could be linked to the presence of active galactic nuclei (AGN), which would be the best signs of this. One team may have identified a NAG located in the galaxy UHZ1, which is believed to be about 450 million years old after the Big Bang, while the JWST Advanced Deep Extragalactic Survey (JADES) may have discovered a NAG in the galaxy GN-z11, which is believed to be about 430 million years old after the Big Bang. The CEERS study also identified 11 distant candidate galaxies that existed when the Universe was only 470 million to 675 million years old. 

\subsubsection{Some Massive Anisotropy Phenomena}

In principle, the Universe is assumed to be globally isotropic and homogeneous on a large scale, despite the few temperature fluctuations detected in the cosmic microwave background, which are at the origin of the large galactic and supragalactic formations we know. However, advances in the spectroscopy of distant galaxies have revealed, over the past twenty years, numerous deviations from this cosmological principle. The discovery of immense groups of quasars such as U127 or Huge LQG constituted a first departure from this, soon followed by the discovery of "large walls", such as the large RGB wall of Hercules-Corona Borealis (HCB) in 2013 or the "large BOSS wall" (BGW) in 2016. The first occupies a volume of at least $9.6 \times 7 \times 0.9$ billion light-years and would extend up to 3000 parsecs (about 10 billion light-years), that is to say over nearly a quarter of the radius of the observable universe (estimated at 46.5 billion light-years). The large BOSS (Baryon Oscillation Spectroscopic Survey) wall, meanwhile, is thought to be located between 4.5 and 6.5 billion light-years away ($0.43 < z < 0.71$) and extends for nearly 1 billion light-years.
In 2021, another form of anisotropy was spotted with the chance discovery of a "Giant Arc" at $z \sim 0.8$, extending over nearly 1 billion light-years.

In the 2020s, another form of anisotropy was identified with the chance discovery of a "Giant Arc" at $z \sim 0.8$, extending over $\sim 1 Gpc$ and appearing to be nearly symmetrical in the sky (see \cite{Lop1}). More recently, the same team demonstrated the existence of a "Big Ring" with an overdensity constituting a significant departure from other structures (see \cite{Lop2}), these two "constructions" located at a distance of approximately 9.2 billion light-years from Earth, yet separated by an angle of only $12^\circ$. The origin of these mega-structures remains, for the time being, unknown and unexplained within the framework of the Standard Model.

\subsection{Other observations}

Still other observations could call the previous model into question.

\subsubsection{The question of the age of the universe}

Despite good corroboration with a number of observations related to Hubble's law and the calculation of the distances of certain stars such as Cepheids, the Universe could be older than previously thought.

Deep-space observations by the James Webb Telescope (JWST) have revealed that the structure and masses of galaxies in the very early Universe, linked to high Doppler shifts ($z \sim 15$) and existing at $\sim$ 0.3 Gyr\footnote{A gigayear (or Gyr for short) is equivalent to one billion years.} after the Big Bang, could be as evolved as those of galaxies existing since $\sim$ 10 Gyr. The JWST results are therefore in strong tension with the $\Lambda$CDM cosmological model. One possible answer could be that light "fatigue" is disturbing the data. But, although tired light (TL) models have been shown to be consistent with the JWST galaxy size angular data, they cannot satisfactorily explain the isotropy of the cosmic microwave background (CMB) observations nor fit the supernova distance modulus well to the Doppler shift data. Some scientists have then developed hybrid models that incorporate the concept of tired light into an expanding universe -- with no more success. Thus, a hybrid $\Lambda$CDM model fits well to the type 1a supernova data but not to the JWST observations. The authors of a recent study (see \cite{Gup}), however, produced a model with covariant coupling constants (CCC), from the modified FLRW metric and the resulting Einstein-Friedmann equations, as well as a hybrid CCC + TL model. These models apparently fit the data admirably, and the CCC + TL model is consistent with the JWST observations. This requires extending the age of the Universe to 26.7 Gyr with 5.8 Gyr at $z = 10$ and 3.5 Gyr at $z = 20$, which provides sufficient time for the formation of massive galaxies. If confirmed, such a model would solve the "impossible first galaxy" problem without requiring the existence of primordial black hole seeds or a modified power spectrum, with rapid formation of massive Population III stars and super-Eddington-type accretion rates. This CCC model could be deduced as an extension of the $\Lambda$CDM model with a dynamical cosmological constant.

\subsubsection{The Initial Singularity Problem}

The "big bang" theory, since its earliest formulations, has given rise to the idea of an initial singularity from which the observable universe results. The problem is that physicists do not particularly like singularities, points at which physics generally gives up. One of the motivations, in the past, for Hoyle's steady-state theory was precisely to avoid Lema\^{i}tre's thesis of the primitive atom and to maintain the idea of an eternal universe. Another way to escape the "Big Bang" models was the Hartle-Hawking construction of a universe without boundary (see Fig. 4).

\begin{figure}[h] % figure placement: here, top, bottom, or page
\vspace{-1\baselineskip}
\hspace{8\baselineskip}
\includegraphics[width=5in]{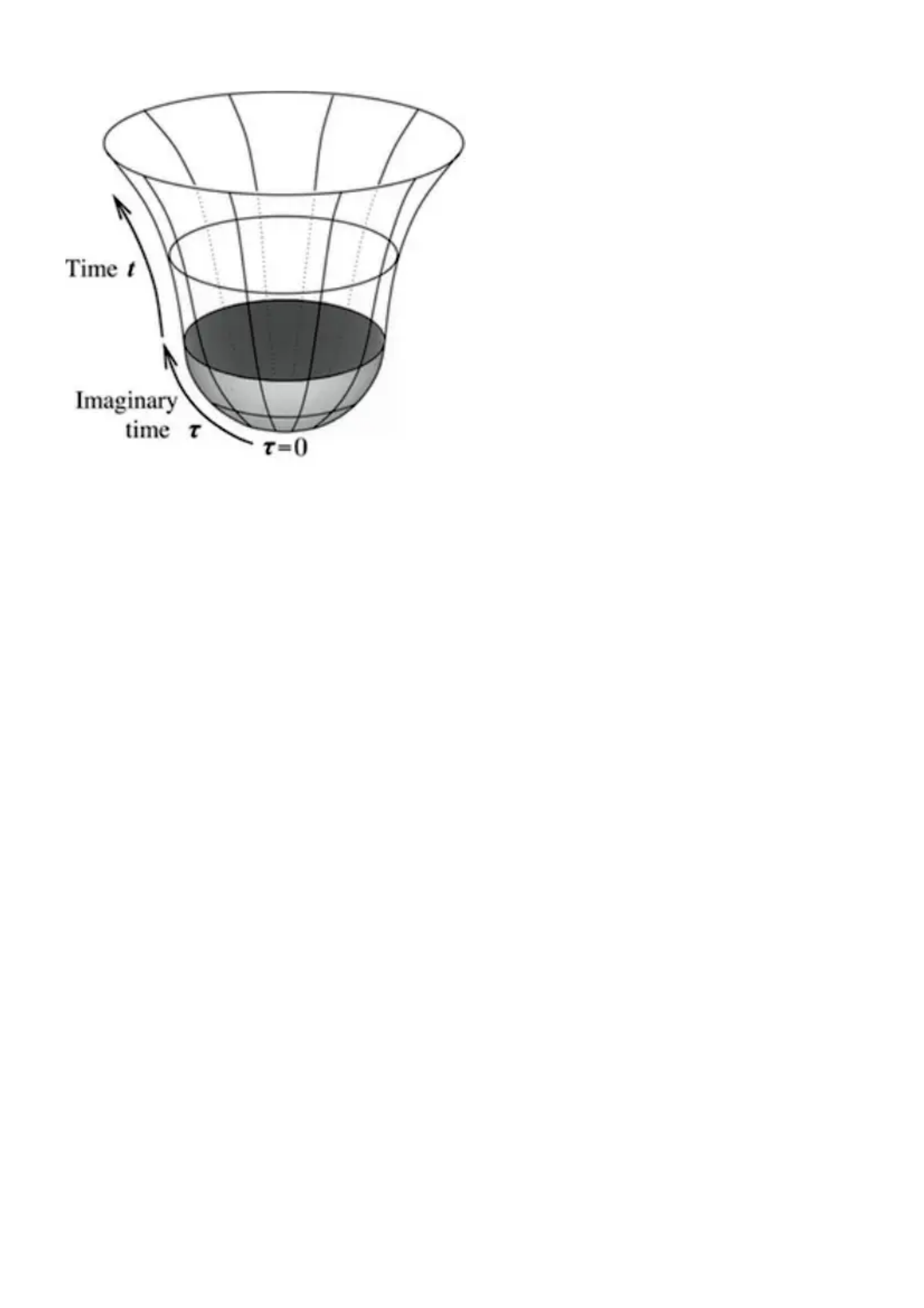}
\vspace{-24\baselineskip}
\caption{The Hartle-Hawking Universe Without Boundaries}
\label{fig: conj11}
\end{figure}

Derived from the theory of Euclidean quantum gravity proposed by Stephen Hawking and James Hartle in the early 1980s, the Hartle-Hawking model describes a universe without boundaries, which does not begin with a singularity of space and time. This model is in fact a wave function of the Universe calculated from the Feynman path integral\footnote{This integral is a reformulation of quantum mechanics leading to the sum of the probability amplitudes for a physical system describing histories, possible paths in space-time between two states of this system. We have commented on Feynman's approach in \cite{Par1}).}.More precisely, it is a hypothetical vector in the Hilbert space of a theory of quantum gravity, which describes this wave function. It is a functional of the metric tensor defined on the compact surface of dimension $D - 1$ (the Universe), where $D$ is the dimension of space-time (see \cite{Hart} and \cite{Haw1}). The precise form of the Hartle-Hawking model is the path integral over all $D$-dimensional geometries possessing the required induced metric at their boundary. According to Hartle-Hawking theory, time diverged from the initial three dimensions after the Planck era. A new direction of time is created before the Big Bang, an imaginary time similar to a complex number.

This model, which assumes the introduction of an imaginary time, might seem artificial, but it is not without motivation. The introduction of an imaginary time is a common practice in physics, allowing for many calculations in quantum field theory. However, this introduction is generally considered merely a computational trick in a common physical theory. The situation is quite different near time zero in a cosmological model. The idea is that, in such a neighborhood, the nature of time is likely to change, so that spacetime will not end with a singular point marking the boundary of time and space. Imaginary time would reflect the fact that space and time are no longer distinct.

The "initial state" of the universe -- if one can still express it that way (but it's obviously an abuse of language) -- would be rather similar to the geometry of the surface of a sphere, except that the sphere would have four dimensions instead of two. Now, just as it is possible to travel completely around the surface of the Earth without ever encountering an edge, one will not encounter a boundary at the origin of the Universe. Thus, the "Big Bang" would no more be the beginning of the space-time associated with the Universe than the North Pole would be, on Earth, the beginning of the Earth's surface.

It has been shown that this wave function of the Universe -- to which the Hartle-Hawking solution corresponds -- satisfies the Wheeler-DeWitt equation, from which all models of quantum cosmology are based. It also has the advantage of escaping the singularity of the "Big Bang" while maintaining the expansion of the universe, thus avoiding the criticisms that could be made against the steady state theory.

\subsubsection{The Possible Existence of Previous Universes}

Another way to avoid the initial singularity and the difficult question of the pre-Big Bang is to assume a universe (or universes) before the Universe.

Let us begin with the theory of an indefinite multiplicity of successive universes. Recall that the Greeks, with the theory of the Eternal Return, based on a physical system where energy expenditures were restored over a long period of time, had considered the idea of a world periodically returning to its initial state. This was the theory of Anaximander of Miletus, to which Anaximenes, Parmenides, and Empedocles of Agrigento may have still adhered (see \cite{Sec} and \cite{Mug}). As Louis S\'{e}chan once commented, "in the unlimited course of time, the world as we see it repeats countless identical worlds of the past, and it is the model that an infinite number of future worlds will reproduce in the future" (Sec, 360).

\paragraph{Tolman's Early Research}
The modern version of this ancient myth -- revised and corrected by relativistic cosmology -- is the theory that the Universe could experience an incessant series of phases of expansion and contraction.

This possibility was explored as early as the 1930s by the American Richard Tolman (see \cite{Tol1} and \cite{Tol2}, 436-438). He was the first modern theorist to imagine an infinite number of cycles for the Universe, which, like the phoenix periodically burning itself, would also periodically rise from its ashes. But such a perspective poses many problems, particularly with regard to the second law of thermodynamics, according to which entropy can only increase in a closed system. To respect this fundamental law -- as Richard Tolman has shown -- the radius of the Universe and the duration of each cycle would also have to increase over time, like a ball bouncing higher and higher, without anyone knowing how long it has been bouncing or how far it will go. Moreover, since 1998, we have known that the expansion of the Universe is accelerating under the effect of the famous "dark energy," which poses new problems. In this context, the expansion, resulting from a dynamic energy field evolving over time, would then have to transform to trigger a contraction. The sequence of hypotheses associated with this rebound theory therefore seems highly speculative.

\paragraph{Penrose Theory}
A variant of these cyclic theories, due to Roger Penrose, does not assume successive expansions and contractions, but rather a simple passage from one universe to another, which retains the trace of the previous one (see Fig. 5).

\begin{figure}[h] % figure placement: here, top, bottom, or page
\vspace{-0.5\baselineskip}
\hspace{2\baselineskip}
\includegraphics[width=6in]{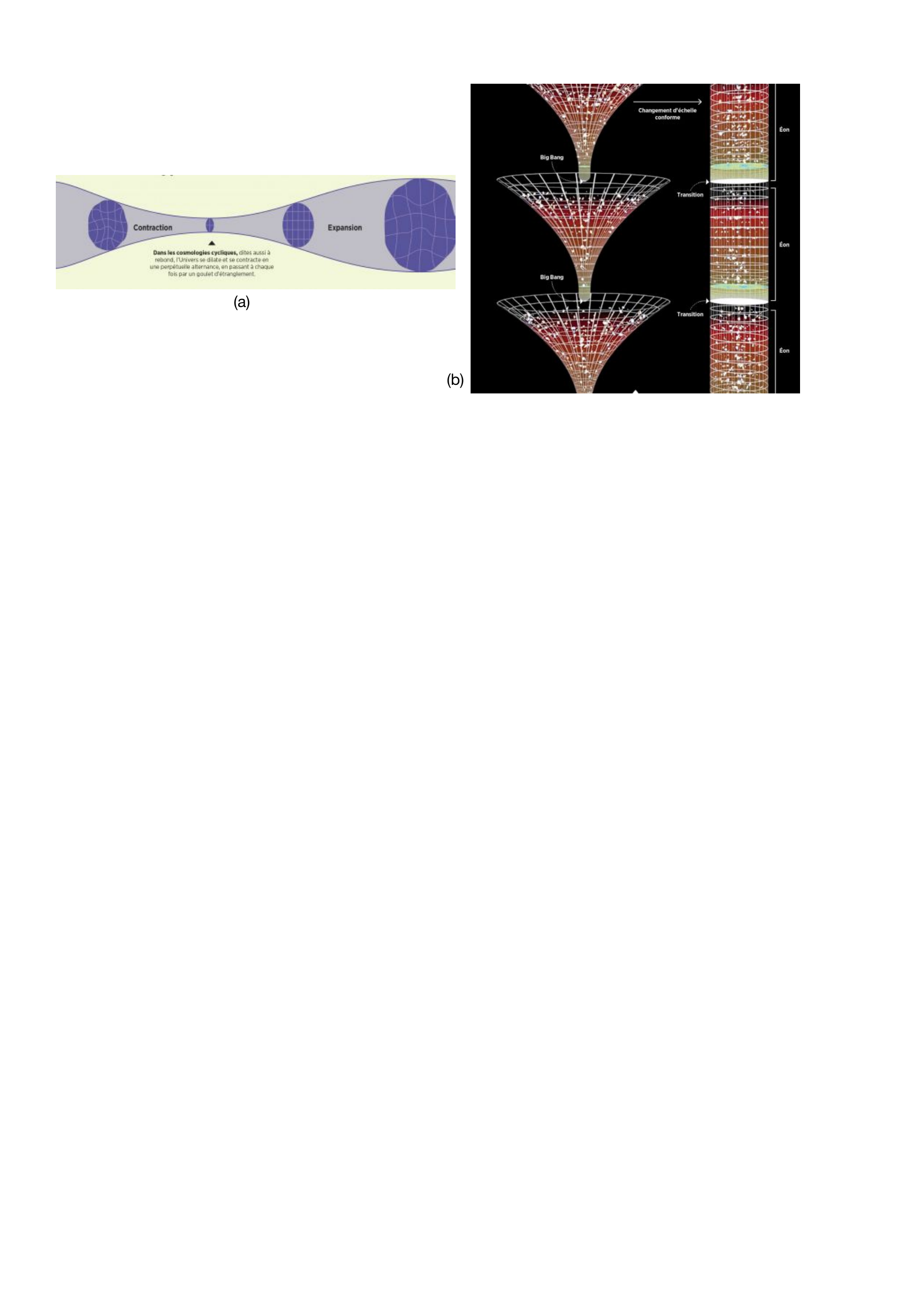}
\vspace{-30\baselineskip}
\caption{Classical (a) and conformal (b) cyclic cosmology}
\label{fig: conj11}
\end{figure}

Penrose's theory originates in a 2006 article (see \cite{Pen1} where the author develops the idea of a conformal cyclic cosmology (CCC)\footnote{In the sense of the "conformal group" which, in geometry, conserves not distances, but angles.}. The starting point of his reflection is the existence of the quasi-homogeneous diffuse background of the Universe at $2.7^\circ$ K, which can be assimilated to a thermal equilibrium, therefore to a maximum entropy, whereas the initial entropy of the "Big Bang" can only be, in principle, very low. Although, in general, there is no clear geometric measure of the entropy in a gravitational field in general relativity, Penrose proposes to associate the situation of the "Big Bang" with a non-activation of the gravitational degrees of freedom. This amounts to saying that the Weyl curvature, that is, the part without a trace of the Riemann tensor associated with spacetime and which describes the gravitational degrees of freedom\footnote{Noting respectively $R_{abcd}, R_{ab}, R$ and $g_{ab}$ the Riemann tensor, the Ricci tensor, the scalar curvature and the metric tensor, the Weyl tensor $C_{abcd}$ is written:
\[
C_{abcd}=R_{abcd}-{\frac {1}{n-2}}\left(R_{ac}g_{bd}-R_{ad}g_{bc}+R_{bd}g_{ac}-R_{bc}g_{ad}\right)+{\frac {1}{(n-1)(n-2)}}\left(g_{ac}g_{bd}-g_{ad}g_{bc}\right)R,
\]
where $n$ is the dimension of the space considered.

In particular, in general relativity, where we almost exclusively consider 4-dimensional spacetimes, we have (see \cite{Din}):
\[
C_{abcd}=R_{abcd}-{\frac {1}{2}}\left(R_{ac}g_{bd}-R_{ad}g_{bc}+R_{bd}g_{ac}-R_{bc}g_{ad}\right)+{\frac {1}{6}}\left(g_{ac}g_{bd}-g_{ad}g_{bc}\right)R.
\]}, must be such that $C_{abcd}=0$ at the initial singularity, while the final spacetime singularities (as occurs inside black holes) are unconstrained.

Using a formulation by mathematician-physicist Paul Tod (see \cite{Tod}) which assumes that an initial space-time singularity can always be represented as a past and smooth boundary in a conformal geometry, Penrose, following Tod, applies this formulation to Weyl curvature. In this case, knowing that the structure of spacetime is invariant under a rescaling of the metric of the type
\[
g_{ab}\to \hat{g}_{ab} \Omega^2 g_{ab},
\]
where $\Omega$ is a smooth positive scalar field, the idea is therefore to "stretch" the metric by a conformal factor $\Omega$, which becomes infinite at the location of the Big Bang singularity. This gives us a metric $\hat{g}_{ab}$ that extends over this singularity and completely smooths the boundary.

Initially, Tod's proposal of a conformal "spacetime" before the Big Bang was merely a mathematical fiction intended to formulate the Weyl curvature hypothesis in a mathematically neat way. Penrose's proposal is to take this mathematical fiction seriously, that is, to consider it as something physically real, in other words, to associate a physical situation with it. In this case, this way of seeing allows him to connect the final state of infinite expansion of the Universe with the initial state of the "Big Bang" without the constraint linked to the singularity which prohibits this kind of connection, which allows him to make physical predictions. The only constraint is that the model adequately describes the state of the elementary particles in the final state of the universe and that the equations characterizing them are invariant under conformal transformations, which clearly means that the particles in question must be massless. This suggests that, in the distant future, matter should lose its properties, the massive particles eliminating or neutralizing each other. In a final phase at very high temperature, the influence of the Higgs field (which gives its mass to particles) will cancel the residue of massive particles. Only photons and gravitons will remain. This conformal cyclic universe is therefore a succession of leaps, each universe, of a higher entropy than the previous one, also keeping its trace within it.
	
Penrose updated his theory in 2010 (see \cite{Gur1}), following data provided by the WMAP (Wilkinson Microwave Anisotropy Probe) satellite which provided a first image of the diffuse background at $2.7^\circ$ K, then in 2015, in an article written in collaboration with the Armenian physicist-mathematician Vahe G. Gurzadyan (see \cite{Gur2}). In the meantime, Penrose had developed his cosmological conceptions more extensively in a popular book (see \cite{Pen2} where the world is broken down into an indefinite succession of phases called {\it aeons}, perhaps in memory of Gnosis, and corresponding to particular expanding universes separated by transition phases (see \cite{Pen2}, 141). The expansion of each universe is transformed at the end of the cycle into a reabsorption into its initial state, but the conformal representation makes it possible to smooth out the singularity\footnote{This theory, like all cyclical theories in general, obviously brings to mind the Greek myth where a giant serpent containing the world, the Ouroboros, bites its tail, drawing this elementary figure of the circle in which all reality is summed up (see Guillaume Duprat's article on the forms and structures of the universe in ancient and oral traditions in \cite{Bar2}, 171).}. Finally, in 2018 (latest version deposited on {\it ArXiv} in 2022), Penrose, returning to "hard mathematics," revised his predictions once again, in collaboration with mathematician Daniel An of the State University of New York and theoretical physicist Krysztof Meissner of the University of Warsaw (see \cite{An}). In this last article, the authors suggest that the diffuse background at $2.7^\circ$ K, of which we now have good images, would contain the physical traces of phantom black holes (Hawking points) representing each time the remnants of dead black holes having merged into a single singularity marking the end of a universe. The gravitons and photons produced by such black holes would have each time created the conditions for a new "Big Bang" and thus produced a new Universe (the black hole thus leading to a white hole). The Hawking points would appear as circles of light on the map of the diffuse background (see Fig. 6).

\begin{figure}[h] % figure placement: here, top, bottom, or page
\vspace{-0.5\baselineskip}
\hspace{9\baselineskip}
\includegraphics[width=6in]{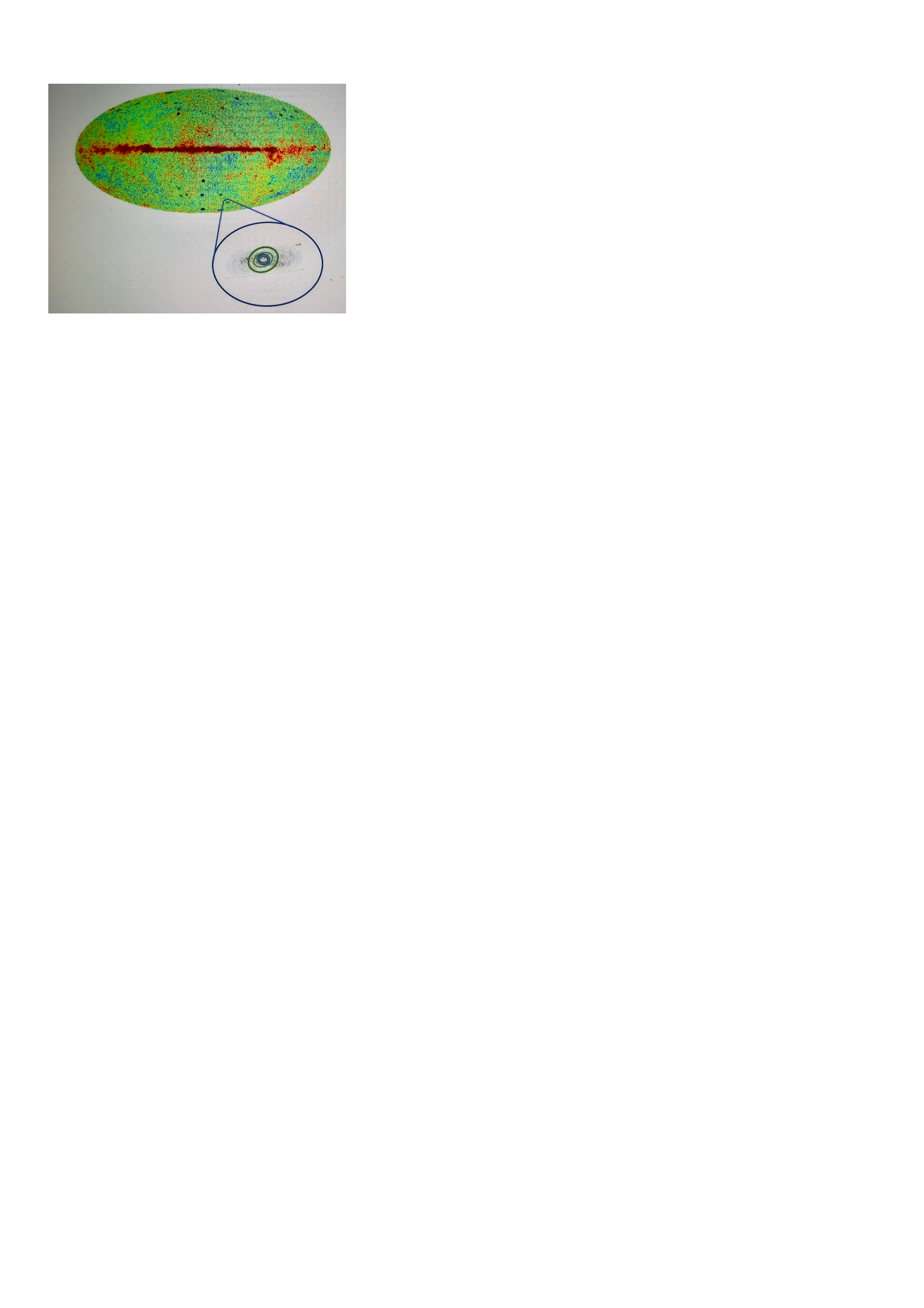}
\vspace{-33\baselineskip}
\caption{Alleged traces of ancient universes}
\label{fig: conj11}
\end{figure}

Despite its potential appeal, the theory does not yet seem to have convinced the cosmology community. Contrary to the authors' assertions, the latest satellite data (the Planck mission, in particular) cannot be relied upon to validate the presumed traces of earlier universes. Indeed, the "circles" associated with the "Hawking points" identified by Penrose and his colleagues could just as easily be considered "noise". It has been shown, for example, that other figures such as curved triangles can also be detected in the diffuse background at $2.7^\circ$ K. Since, moreover, there is no proof that the phantom black holes of previous universes obey the same laws as those known in the present universe, the whole theory remains too speculative to gain acceptance. But we can still give it a chance by assuming that the postulated traces remain below the detection threshold of our instruments for the time being.

\paragraph{Cycles in the Framework of String Theory}

Other cyclical cosmologies (or those with bouncing universes) have been developed in the last twenty years (see \cite{Nov} and, more recently, \cite{Nev}), some of which are related to string theory.

Indeed, the great pioneer of this theory, Gabriele Veneziano, had addressed the subject in the early 2000s. The idea that, in certain string models (pre-Big Bang models or ekpyrotic models), scenarios where time has neither beginning nor end could have left traces in the cosmic microwave background was in the air (see \cite{Gas1}).

Veneziano, like all cosmologists, was already questioning the extreme homogeneity of the diffuse background radiation at 2.7 K at this time. Given its identical consistency in all directions, this implies that galaxies located in opposite directions, and therefore separated by about 25 billion light-years, are roughly the same, even though the regions in which they are located have never come into contact. Two explanations were possible for this anomaly: "Either the universe was, in its earliest moments, much smaller than classical cosmology assumes, or it is much older. In either case, two distant parts of the sky before the emission of cosmic radiation could have interacted" (see \cite{Ven}, 42). The first solution is that of the supporters of inflation, a theory proposed by Englert, Brout and Gunzig in 1978, then Starobinsky in 1979, then simplified by Alan Guth in 1981 (see \cite{Pet}, 427). But difficulties remain, inflation being based on totally exotic physics, with a mysterious "inflaton"\footnote{Recall that the "inflaton" is the name given to a hypothetical scalar field (that is to say a place in space where a force is exerted that is not oriented, unlike vector fields). This scalar field would resemble that associated with the Brout-Englert-Higgs boson. It is this inflaton which would be responsible for this phase of extremely rapid accelerated expansion at the very beginning of the universe that is inflation.}. whose nature is unknown. The second way to solve the problem -- common to many alternative theories to the "Big Bang" -- is to assume that the Universe is much older than expected. In this case, that is, if a long era preceded the current epoch, the universe has indeed had time to homogenize. This second scenario, as previously assumed by Penrose and the other theorists mentioned above, has the advantage of eliminating the singularity, which prohibits the application of zero-point relativity. This is then replaced by a quantum theory of gravitation, with two promising approaches: loop quantum gravity and string theory. It is the latter that Veneziano is concerned with. It highlights particular properties of strings, well adapted to quantum laws: size of the order of $10^{-34}$ meter, vibrational energy corresponding to the masses of the particles and adapted movements, equations consistent with quantum physics if space has 9 dimensions and if the 6 additional ones are wrapped over small distances, constants describing the intensity of the fundamental forces appearing in the form of fields with evolving values as a function of time, existence of new interesting symmetries, not existing in the theory of relativity (see \cite{Ven} 43-44),

Precisely, among the symmetries manifested by strings, one of them, T-duality, reveals that a universe where the scale factor is very small is equivalent to a universe where the scale factor would be very large. This T-duality thus sets limits to certain physical quantities and notably prevents the radius of curvature of the universe from decreasing to zero and causing the problematic singularity. It causes the collapse to rebound, which becomes a new expansion. From then on, once the singularity has disappeared, nothing prevents us from assuming that the Universe existed before the "Big Bang". It can then take two forms: in the pre-Big Bang scenario, it is a mirror universe, so that the total Universe extends eternally into the future as in the past; in the ekpyrotic scenario, two branes meet and rebound before falling back on each other. Apparently, we could eventually know which model prevails over the other. The pre-Big Bang and ekpyrotic scenarios predict more high-frequency gravitational waves and fewer low-frequency ones than the inflationary model. It is therefore possible that this question may eventually be resolved (see \cite{Ven}, 45-47). Experience speaks for itself.

In a technical article, the theory, reformulated in Hamiltonian language, characterizes a broad class of non-singular and non-perturbative pre-Big Bang scenarios, a kind of interpolation between an initial accelerated expansion (string frame) of low energy and a final slowed expansion phase (string frame and Einstein frame). In all cases, these solutions must necessarily include, just around the rebound, a very short contraction phase (described in the string frame) (see \cite{Gas2}).

Ultimately, all these models of the universe, as Aur\'{e}lien Barrau demonstrated in a book in which he lists them (see \cite{Bar1}), are situated within the context of a theory of the multiverse, which is sometimes considered in a spatial form (where universes coexist), and sometimes in a temporal form (where they succeed one another).

\subsubsection{WMAP and the Shape of the Universe}

Another interpretation of the anomalies in the cosmic microwave background, and in particular the "circles" that can be identified there and on which cyclic theories like Penrose base their arguments, could be linked to the shape of the universe. The theory of relativity left the topology of the universe undetermined. Since the beginning of the 21st century, however, certain observations from telescopes have shown that the shape of our universe is approximately flat. This allows us to envisage different types of space of this nature.

An analysis of data from the WMAP artificial satellite by Jeffrey Weeks, Jean-Pierre Luminet, and their collaborators (see \cite{Lum}) had already suggested a Universe whose shape would be that of a Poincar\'{e} dodecahedral space. Jean-Pierre Luminet expressed the idea that the Universe could be of finite spatial extension, but without boundary, what he called in his popular books and articles a "crumpled universe." The more accurate scientific expression is rather "universe with multiply connected topology" or "not simply connected."

More recently, a German team from the University of Ulm, led by Frank Steiner (see \cite{Ste}), confirmed that the shape of the spectrum of temperature fluctuations in the cosmic background radiation in a finite-sized Universe with multiply connected topology was compatible with the WMAP data. But unlike the French researchers who retained the shape of the dodecahedron, Steiner and his team used a simpler model of a crumpled universe: a three-dimensional torus with a volume equal to 5 $\times 10^3$ Gpc. This object of zero curvature satisfies the theorems of Euclidean geometry. On the other hand, like the Poincar\'{e} model, it allows the appearance of correlation circles in the fossil radiation as well as the multiplication of ghost images of galaxies, suggesting that the universe is infinite.

In the cosmic microwave background radiation spectrum, the long-wavelength deficit compared to the predictions of the $\Lambda$CDM model (see left of Fig. 3) could be explained as a limit to the size of temperature fluctuations in the cosmic microwave background radiation, a limit imposed by the existence of a closed, and therefore finite, universe. (Apparently, it seems that the $\Lambda$CDM model with an infinite universe fails to recover this wavelength deficit.)

Another argument in favor of the German cosmologists' theory is that, as Andrei Linde had already shown, in quantum cosmology, the birth of a torus-shaped universe is much more likely than that of a finite, spherical or, conversely, flat but infinite universe. Although this quantum cosmology remains hypothetical, it presents the beginning of the universe as a quantum transition by tunneling between a physical state dominated by quantum gravity, in which neither space nor time have their usual properties (or even cease to exist, except in virtual form with an imaginary time), and a classical state associated with the space-time geometry we know. In this context, the emergence of a compact and flat geometry then seems more likely, as Linde argued in his work.

But we will probably have to wait a few years to confirm or deny what remains, for the moment, speculation.

\subsubsection{The long-term future of the Universe and the physics of eternity}

Within a cosmological conception where the Universe has a beginning, it must also have an end. Speculation about the distant future of the observable universe has flourished for about forty years, notably under the influence of American physicists such as Tipler (see \cite{Tip}), Deutsch (see \cite{Deu}), and Redford (see \cite{Red}). As we reported in one of our publications (see \cite{Par2}), a fairly optimistic vision of the future seemed to emerge from this long-term outlook, even if Dyson's initial estimates (see Dys) for an eternally expanding universe had since had to be revised downwards (see \cite{Ega}).

In the 2010s, the evaporation of the last black holes was predicted to take approximately $10^{100}$ years (more precisely, $10^{80}$ years for stellar black holes and $10^{110}$ years for supermassive black holes).

More recent studies seem to extend the life of the universe well beyond that. Following the reflections of Fred Adams and Greg Laughlin (see \cite{Ada}), a recent article by Matt Caplan (see \cite{Cap}) shows that black holes are not the last objects to survive. It would be the explosion of white dwarfs, hundreds of trillions of years from now, that could constitute the final cosmic fireworks display, before the universe sinks into total darkness, devoid of all life.

But this final cooling is only one possibility among others. If, on the contrary, all the remaining matter concentrates to the point that the Universe collapses in on itself, it is not impossible that this final apocalypse is in fact only the beginnings of a new Big Bang.

\section{Philosophical and Theological Consequences}

Cosmology, in principle, is a science in its own right, moreover, highly mathematized since the years 1915-1920, like Albert Einstein's general relativity. Nevertheless, some cosmologists believe that it has not completely freed itself -- and perhaps will never be able to do so entirely --  from a cultural environment that weighs heavily on it. As Marc Lachi\`{e}ze-Rey writes, "studying fundamental problems that call into question the nature of our universe, the question of its origin and its future, as well as the place occupied by man in it, cosmology shares its centers of interest with extra-scientific disciplines, metaphysics, theology, philosophy... It is not always easy to establish a barrier between what is scientific and what is metaphysics, nor to recognize the influences of the different myths that haunt us, or of our socio-cultural heritage. In any case, it is important to keep in mind that scientific cosmology is based on metaphysical principles." (see \cite{Lac}, 3).

However, it is one thing to detect an unintentional mixture of philosophy and science in cosmology, and quite another to assess how modern cosmology may have changed our representation of the world. Yet, on this point, few philosophers or theologians have been willing to draw any lessons from contemporary cosmological speculations.

\subsection{Concordism and its Refutation}

The greatest error, from this point of view, was certainly committed by Pius XII, a great defender of "concordism," this attempt to reconcile science and the Christian faith. The Pope relied on a text by Edmund Wittaker, a papal academic, suggesting that the initial moment of the age of the world, to which cosmological research now goes back, was in some way "the ultimate limit of science" and that, therefore, we could perhaps "refer to it as creation" (see \cite{Whi}, 118-119). Pius XII also allowed himself to purely and simply identify the "Big Bang" with divine creation: "It seems, in truth, that science today, going back millions of centuries in one go, has succeeded in making itself the witness of that initial "Fiat Lux", of that instant when an ocean of light and radiation arose from nothingness, along with matter, while the particles of the chemical elements separated and assembled into millions of galaxies... It has considerably broadened and deepened the bases of experience on which the argument is based and from which one concludes the existence of an {\it Ens a se}, immutable by nature... Thus, creation in time; and, for this, a Creator and, consequently, God! Here then - even if implicit and imperfect - is the word that We asked of science and that the present human generation expects from it." (see \cite{Pie}, 1537 sq). As we know, it took all the weight and influence of Abb\'{e} Lema\^{i}tre to rectify the papal position, whose speech of September 7, 1952, before the International Astronomical Union, then referred only to "the cosmic processes that took place on the first morning of Creation," without further clarification (see \cite{Rob}, 120-121).

Given the uncertainties of science -- which is obviously part of history -- and what Jacques Merleau-Ponty called this "dialectic of origin" (see \cite{Mer}, 337), which seems to extend without resolving the Kantian antinomies of the Critique of Pure Reason (see \cite{Kan1}, 338 sq), one might already consider Lema\^{i}tre's refusal to bring together the discoveries of science and the assertions of religion as prudent. But there are still deeper arguments that can be invoked, which we will detail later and which relate to the notion of "creation" (see section 3.6).

\subsection{An Abusive Criticism of Materialism}

Another error, in our opinion, is made when one believes that one can infer from the advances of science the rejection of certain philosophical conceptions, moreover often described in a crude if not caricatured manner. Claude Tresmontant, in the 1970s, unfortunately followed this path. In the first chapter of his book {\it Sciences of the Universe and Metaphysical Problems} (see \cite{Tre1}), which deals with cosmology, the author, under the pretext of combating the ignorance of the philosophers of his time in scientific matters, undertakes to draw ontological consequences from modern cosmology. The latter, which shows the existence of a dynamic, evolving universe, opposed to the static system of Antiquity, which knew "neither growth nor decline, neither beginning nor end" therefore teaches "just the opposite of what ancient Greek philosophy taught." One can, of course, deplore with the author that many 20th-century philosophers -- including Heidegger and Sartre -- have been able to maintain that philosophy was unrelated to science. But one cannot follow him in the reasoning he uses in opposing modern cosmology to ancient philosophy and to what he claims to be its identification of the universe with the totality of being, a thesis which is, according to him, that of Parmenides as well as Aristotle or even Heraclitus. Since being cannot have begun, it follows that the universe is eternal, an idea defended, according to the author, by all materialists. For Tresmontant, therefore, modern cosmology, which he assumes has definitively settled the question of origin by positing that the universe began, refutes materialism just as it also refutes the theories of eternal return, from Heraclitus to Nietzsche -- other eternalist theories. According to Tresmontant, a last square of scientists -- "the old guard" -- "still defends, against all the data of astrophysics, the hypothesis of an eternal universe -- because they know: if the thesis of a beginning of the universe prevails, it is the end of materialism (...)". For, for what he calls “atheistic materialism” to be true, the universe must be imperishable, and it must be Being, the only Being, all of Being. To this tradition of Parmenides and the Greek thinkers, Tresmontant opposes the thought "which goes back to the ancient Hebrews" and according to which the universe is some being, but not Being taken absolutely, which would explain that it could have begun (Genesis) and that it could end (Revelation). However, Tresmontant must recognize that, if by chance cosmology opted for a universe without beginning or end - an eternal universe, in short -, such an observation would not necessarily prove materialism. This is therefore an implicit admission that, contrary to the initial assertions, no ontological conclusion can truly be drawn from the new cosmology - which does not mean, of course, that a philosopher should not study it. However, one can only contest the idea that the existence of the "Big Bang" would refute materialism, the universe, in its beginning (Lema\^{i}tre's primitive atom or current quantum foam) as in its later developments (clusters, nebulae, galaxies, etc.) being all that is "material". Conversely, one does not see why certain scientists (Bondi, Gold, Hoyle) should cling to the idea of an eternal universe for the sole purpose of defending materialism, since the eternity of the Universe will not succeed in proving it. The whole reasoning therefore seems biased, as is also proven by the rest of the book, which aims to show that the modern conception of matter, like that of living beings, no longer allows us to rely on chance and to dispense with the intervention of an organizing intelligence, which is obviously the initial presupposition, which continues to run through all the chapters -- including those on man, knowledge, causality or evil.

A more recent book by Tresmontant (see \cite{Tre2}) does not seem much more convincing, despite reasoning that is understandable but which definitely deviates from science: "A very large number of astrophysicists believe that this Universe in which we are has a beginning. But to say that the Universe has a beginning is not yet to say that it has been created. To establish that it has been created, and that it is currently being created, one must proceed to a rational analysis, which is a metaphysical analysis. To note the beginning of a being is not yet to have discovered that this being has been created. For the idea of creation strongly implies two terms. One of the two terms is the created being. The other term is the creative being. As long as we have not discovered the existence of the Creator, we have not discovered creation, which is a relationship between the Creator and the created." (see \cite{Tre2}, 179). We are no longer talking here about a Being pre-existing the being of the Universe (the risk was undoubtedly great that the existence of this Being would eliminate creation). What is now envisaged is the idea of a continued creation, but not in the Cartesian sense, rather the progressive creation of a new humanity via revelation. Tresmontant then abandons to the sciences the material past of the Universe, to credit Christianity with a vision of the future: "The past of creation is known by the experimental sciences -- astrophysics, physics, chemistry, biochemistry, fundamental biology, zoology, paleontology, neurophysiology; -- the future of creation is known by revelation" (see \cite{Tre2}, 186). With the appearance of man already comes a kind of creation of a spiritual order, but with Christianity is revealed "the creation of the New Man united with God" (see \cite{Tre2}, 191). Theology, contrary to Tresmontant's initial wishes, has thus completely separated itself from the scientific cosmology on which it initially sought to rely, becoming purely apologetic. It is now a matter of covering the history of the universe with a narrative of another kind, one that draws its origins and all its meaning from Hebrew prophecy.

\subsection{The Decoupling of Science and Faith}

Seen today, that is, almost half a century later, one might think that Tresmontant's words have fallen into disuse. Commenting recently on Gr\'{e}goire Celier's work on {\it Saint Thomas Aquinas and the Possibility of a World Created without Beginning} (see \cite{Cel}), Jean-Fran\c{c}ois Stoffel praised the author not only for the accuracy of his analyses, but for what they signify, in relation to the history of 20th-century cosmology and current research. "By recalling that creation is in no way amenable to science since it applies not to an already existing mobile being, but only to the absolute appearance of being (p. 178 and pp. 182-187), this presentation presents, in our opinion, yet another interest: that of freeing scientific research from ideological considerations which have no place there and which, most of the time, are likely to hinder its natural development. Indeed, when a scientist tries, through his work, to "circumvent" cosmological models that state a natural beginning of the world, believing that this will destroy any claim to a metaphysical creation of this same world, he is wasting his time unnecessarily\footnote{It is likely that the scientist targeted is Stephen Hawking. But, more generally, those who fall under this criticism are the defenders of eternal or cyclical universes, from Hoyle to Penrose.}; similarly, when an apologist takes advantage of these same cosmological models to give more credence to the doctrine of the creation of the world, he is working dangerously" (see \cite{Sto}). If we can rejoice in the last sentence, which definitively condemns Pius XII and concordism, or, in a general way, any instrumentalization of science, we can be more circumspect about the other aspects of this text which, faithful to Thomas Aquinas, seems to support the idea that what cosmological science can say is without consequence on metaphysics or theology, because the idea of creation applies to the appearance of Being, not to that of the universe. We are therefore not very far from Tresmontant because the latter distinguished the being of the Universe from Being in general only to maintain, precisely, the necessity of a Creation in the strong sense and therefore, of course, of a Creator. It would be a matter, however, of demonstrating that what is called "Being", and which is supposed to have no property of "beings", nevertheless possesses an existence different from that of a phantom or an imaginary construction. But this Being, strangely, seems to have become so evanescent over time that certain philosophers -- who make it a sort of idol -- do not hesitate to remove it from God himself, thus developing the idea of a "God without being." Against classical metaphysics and the neo-Thomists, Jean-Luc Marion, following in the footsteps of Schelling -- who defended the idea of God's freedom with regard to his own existence -- did not hesitate to take the plunge: "Does being have a relationship, more than anything else, to God?" -- he asks. Does God have anything to gain from being? Can being -- whatever it is, as long as it is, manifest -- even welcome something of God?" (see \cite{Mar}, 11). If being is then only an idol, God must rather belong to a category other than the ontological in the strong sense, and thus find himself -- which is perhaps more in keeping with Christianity -- identifiable with giving, love, and charity, the Christian being able to approach him, precisely, only through his actions. Obviously, all this no longer has anything to do with cosmology.

\subsection{Towards a Possible Autonomization of Cosmology}

It should be noted, moreover, that scientific cosmology tends to become completely autonomous. While acknowledging that there is nothing in his book "that discredits the possibility that an intelligent agent could have created the universe for any purpose," Leonard Susskind, a great specialist in string theory, admits, like Laplace in the past, that he can do without this hypothesis, or at least doubts its validity: "If there was a moment of creation, it is hidden from our eyes as it is from our telescopes behind the veil of explosive inflation that took place during the prehistory of the big bang. If God exists, he took great care to remain unnoticed" (see \cite{Sus}, 576). Better still: one will find no metaphysical or theological reference in the book by Joseph Silk, one of the great defenders of the initial singularity, which bears precisely this title: {\it The Big Bang} (see \cite{Sil1}). In another of his works, a little longer and a little more recent, (see \cite{Sil2}, 12-13), the author explains, moreover, how we moved from the imprudences of Whittaker and Pius XII to this now autonomous cosmology. Arthur Eddington, it is true, refused to admit the first discontinuity, which he identified with divine intervention\footnote{He would have preferred a less unsightly solution: "Since I cannot avoid introducing this question of a beginning, I have the impression that the most satisfactory theory would be one for which this beginning {\it would not be too unsightly sudden}" (see \cite{Edd}, 72).}; Milne still considered that the system of modeling the universe that cosmology had arrived at was an intelligible system, which had nothing irrational about it, except, precisely, this first instant of creation. But Steven Weinberg, on the other hand, believed that it was "logically possible that there was a beginning, and that time itself had no meaning before it" (see \cite{Wei}, 173). For Weinberg, we therefore only had to get used to an absolute zero of time, just as we had gotten used to the idea of an absolute zero of temperature. In which case, we are definitely moving towards a cosmology completely purged of metaphysical expectations.

\subsection{Hawking's Extraordinary Project}

In the conclusion of the book that made him famous in the eyes of the general public -- {\it A Brief History of Time} (1989 for the French translation) -- Stephen Hawking announced that, if the universe had, as he thought he had proven, neither singularities nor edges, and if it were completely described by the unified theory he proposed, then this must have "profound consequences for the role of God as creator". Hawking reported that Einstein had once asked what choice God had in constructing the universe. Hawking's answer took the following form: "If the proposition no edges'  is correct, he had no freedom to choose the initial conditions. Of course he could still have had the freedom to choose the laws by which the universe obeys. However, this does not represent a wide range of possibilities; there may very well be one or a small number of completely unified theories (...) that are consistent and allow for the existence of structures as complex as human beings capable of investigating the laws of the Universe and asking questions about the nature of God." (see \cite{Haw2}, 221). A little further on, the physicist asked himself again: "Is the unified theory so binding that it ensures its own existence? Or does it need a creator and, if so, does this have other effects on the Universe? and who created it?" (see\cite{Haw2}, 221). Hoping that a day would come when philosophers, scientists, and ordinary people would be able to understand the famous complete theory that explains everything, he assured that at that time, men would know "the mind of God" (last words of the book). This formula was surely chosen deliberately: for the Christian religion, "knowing the thought (or mind) of God" is equivalent to putting oneself in his place: it is obviously human pride par excellence, the original sin of humanity, which gives Hawking's project a totally transgressive air.

Around the same time, although the French translation was a little later (1992), Hawking continued to announce, at the end of a lecture on "the edge of space-time," "important philosophical consequences," notably the possibility of describing the Universe using a mathematical model entirely determined by the laws of science and science alone. Since this model cannot predict everything, both because of the uncertainty principle (which cannot accurately give the value of certain quantities, only their probability distribution) and also because of the complexity of certain equations, which are impossible to solve except in simple cases, "we would therefore be," he asserted, "still very far from omniscience" (see \cite{Haw3}, 118-119).

Hawking's lecture entitled "The Origin of the Universe," which was also roughly contemporary with the writing of his first book, was also more cautious. It reads precisely this: "Is the ultimate unified theory so irresistible that it causes its own existence? Although science can solve the problem of how the universe began, it will not be able to answer the question: why did the universe bother to exist?" (see \cite{Haw4}, 97). Therefore, not only will we not be omniscient, but we will not know the ultimate ends of things, we will not answer the metaphysical question par excellence, we will not be able to know, in the end, why there was something rather than nothing.

In the beginning, however, is nothing for Hawking, and since everything is born from this nothing, it is indeed a creation {\it ex nihilo}. It was during Hawking's dialogue with Penrose on the nature of space and time that this point was clarified, in the context of the explanation of the "no boundary" condition. This hallmark of Hawking's cosmology, which amounts to inserting a Euclidean sphere at the base of the Lorentzian hyperboloid of a de Sitter space, leads the physicist to state the following: "Unlike the case of the creation of a pair of black holes, it is not possible to consider that the de Sitter universe was created from the energy of the field in a pre-existing space. On the contrary: this universe was literally created from nothing. The expression "creation from nothing" had already been used by Friedmann (see \cite{Fri}, 206) in the cosmological context, but probably not with the same force as by Hawking: not from a vacuum, but from nothing, absolutely nothing, because there is nothing outside the universe" nxknnaaa(see \cite{Haw5}, 133-134). One could therefore not be more explicit on the issue.

It was legitimate, after this, for Hawking to address the theological question head-on. He does so in the text 'Is There a Great Architect in the Universe?', published in collaboration with Leonard Mlodinow. While acknowledging the legitimacy of the question, as with other metaphysical questions, which were traditionally the domain of philosophy, the authors assert that, now that philosophy is dead, it is up to scientists to answer it. And as expected, their answer is negative.

"Over the centuries, many, like Aristotle, have believed that the Universe has always been present, thus avoiding the pitfall of its creation. Others, on the contrary, have imagined that it had a beginning, using this argument to prove the existence of God. Understanding that time behaves like space allows us to propose an alternative.
This alternative, dismissing the hackneyed objection that opposes any beginning of the Universe, relies on the laws of physics to explain this creation without recourse to any deity" (see \cite{Haw6}, 166).

Simply, since the origin of the Universe was a quantum event, it must be described, according to Hawking, by the Feynman integral, that is, the sum over all possible paths, but not exactly between two points $A$ and $B$, as for particles. In this case, since the point $A$ does not exist, it is sufficient "to add up all the histories which on the one hand satisfy the boundary-free condition and on the other hand lead to the Universe we know today" (see \cite{Haw6}, 167). Some histories will be more probable than others and the sum could be dominated by a single history starting from the creation of the Universe and culminating in the state we are considering. However, it is clear that different histories will correspond to other possible current states of the Universe, with particular laws of nature depending on these histories each time. When we model particles as vibrating strings, the M-theory that is supposed to describe the nascent Universe necessarily includes a certain number of additional dimensions beyond spacetime (six or seven), which must be thought of as folded in on themselves since they are not visible. These dimensions, which constitute the internal space determining the values of physical quantities such as the charge of the electron or the nature of the fundamental interactions between elementary particles, can be folded in a large number of ways. In an ideal scenario, M-theory would have allowed only a single form for the folded dimensions, or at most a small number of forms from which all but one would have been excluded, ultimately allowing only a single choice for the apparent laws of nature. The reality, unfortunately, is quite different: according to the authors, it seems that one can assign amplitude probabilities to no less than $10^{100}$ different internal spaces and according to Leonard Susskind, this figure could even be as high as $10^{500}$. The landscape of possible universes is therefore incredibly vast and the determination of this being that is the Universe particularly weak: "The fundamental quantities and even the form of the apparent laws of nature turn out to be determined neither by logic nor by a physical principle. The parameters are thus free to adopt all sorts of values and the laws to take any form that leads to a coherent mathematical theory" (see \cite{Haw6}, 177). In this immensity of possible universes, those likely to give rise to the constitution of stable structures and then of living and thinking organisms are particularly rare. A certain number of constants must be particularly well-adjusted.

This has thus answered the question of how the Universe behaves. But the why questions remain: why is there something rather than nothing? Why do we exist? Why this particular set of natural laws and not another? Is this a sign of a "grand design"? Involving a God behind the judicious selection of constants that characterizes our Universe would only postpone the problem, since it would then be necessary to explain who created the creator. "We affirm," the authors write, "that it is possible to answer these questions while remaining within the realm of science and without recourse to any divine being" (see \cite{Haw6}). The Universe we know is a bit like the computer game invented by John Conway called the "Game of Life," in which a set of very simple laws can give rise to complex properties. In our Universe, the analogues of the objects in the "Game of Life" are isolated material bodies subject to sets of laws that each assume a principle of invariance, the conservation of a certain energy that must remain constant over time.

Things could therefore work like this: let's set the energy of the vacuum to zero. Every existing body has positive energy, the energy expended to create it. If this energy were negative, indeed, this body could be created in a kinematic state so that its negative energy would be compensated by the positive energy due to its speed. In this case, nothing would prevent bodies to appear anywhere in space, the vacuum consequently becoming unstable. Since the energy of bodies is positive, such instability cannot occur, the energy of the Universe must remain constant. If the total energy of the universe must always be zero and the creation of a body costs energy, how could an entire universe have been created from nothing? The authors' idea is that gravitation plays a crucial role here. Gravitation being an attractive force, gravitational energy is negative. It counterbalances the positive energy necessary for the creation of matter, but not totally. Thus the gravitational energy of a star is negative, and the more the star is gathered, the more negative this energy is. But, before the latter can exceed the positive energy of the matter of which it is composed, it must collapse on itself, giving rise to a black hole whose energy is again positive. As a result, empty space is stable, stars and black holes cannot appear from nowhere. In On the other hand, an entire universe can: on the scale of the universe, the positive energy of matter can be offset by gravitational energy, which removes any restrictions on the creation of entire universes. Thanks to the law of gravitation, the Universe can therefore spontaneously create itself from nothing. Thus, "spontaneous creation is the reason why something exists rather than nothing, why the Universe exists, why we exist. There is no need to invoke God for him to light the fuse and bring the Universe into being" (see \cite{Haw6}, 219).

\subsection{Cosmology and Theology}

Hawking seems to have achieved what no one thought possible: a creation ex nihilo without God, which even Hubert Reeves, not long ago, still considered impossible. Speaking of multiverse models, the author wrote: "Do these scenarios explain creation? In the strict sense of the term, creation is the transition from "nothing" to "something." Here, we are not starting from nothing... "In the beginning were the laws of physics," the authors implicitly say. Physics always connects something to something else. Why "this": Because "that." Explaining creation would be connecting "something" to "nothing," {\it really} "nothing." Such a feat seems beyond the possibilities of the scientific approach... (see \cite{Ree}, 221).

With Hawking's solution, many metaphysical or theological objections therefore seem to fall away. Since we move from "nothing" to "something," it seems that this is indeed an apparition of being. And the objection that cosmology does not explain creation but only the birth of a world that is only a part of being, or even of a universe that is merely "observable," no longer holds.
Moreover, it is not a (single) universe, but a multitude. The creation that was believed to be unique and mysterious becomes the most banal thing since nothingness--zero energy -- cannot help but produce these billions of billions of bubble universes ad infinitum...

The fact remains that this Hawking cosmology, like all other multiverse cosmologies, remains largely speculative. Many aspects of the universe still elude us, and some will always elude us. Beyond the light horizon, the moment when the universe, having become visible, left us a trace with the cosmic microwave background at 2.7 K, we can still speculatively go back to the Planck era (10$^{-43}$ seconds), which is a sort of absolute horizon that is difficult to cross. But, in any case, we also don't know what has become of the galaxies we see today as they were billions of years ago: the cosmological horizon definitively prevents us from doing so, regardless of the cosmological model we choose.

With the exception of Hawking's -- contested -- attempt, almost all scientific cosmology does not consider itself capable of entirely replacing philosophical or religious discourses on the origin of the world, and considering an eternal or cyclical universe doesn't change much. Moreover, according to the texts of Thomas Aquinas himself, an infinite and eternal universe remains in principle compatible with a "creation," even if this idea nevertheless contravenes the Judeo-Christian doctrine according to which the world has a beginning in space and time.

On the other hand, it seems more questionable to us to make a distinction between "world" and "universe," and continuing to speak of a "creation of the world" as distinct from an "origin of the universe" is untenable. At least in French, the words "world" and "universe" are practically doublets: both have cosmological, historical-geographical, sociological, and even psychological and aesthetic meanings.\footnote{One can hardly suggest that the word "universe" -- this might be its specificity -- has more scientific connotations, given its use in expressions such as: mathematical universe, universe of discourse (linguistics, logic), universe in the sense of "group of elements from which one draws a sample" (statistics), and finally, a set that can be equipped with a law of probability (mathematics). In the same vein, one might also note that the opposition world/universe has been used to distinguish the representation of space associated with Antiquity from the modern representation of space resulting from the Renaissance and the Classical Age (see \cite{Koy2} and \cite{Koy1}). But that is all that can be said about this difference between "world" and "universe".}. Certainly, scientific cosmology is limited to physics, but the physical universe is the basis of all others. What stuff would a world be made of, moreover, if it were not, at the outset, physical? As for Tresmontant's idea that the universe would be only one being among others, the theory of multiverses responds to it. This is precisely the Being beyond being, and there is no need for a Deus ex machina to explain its presence.

\subsection{The so-called "creatio ex nihilo"}

Modern cosmologists, from Friedmann to Hawking, have gone to great lengths to understand how the universe could have arisen from nothing.
Now, this approach may, in fact, be pointless, but not for the reasons given by Stoffel. A careful study of the biblical texts shows that the idea of creation from nothing, anchored in minds influenced by Christian culture, is in no way a Jewish idea.\footnote{It is well known that the Old Testament does not mention creation from nothing (see \cite{Sch}).} Only two texts in the Bible are at the origin of this legend, the commonly given translation of which is highly questionable.

The first of these texts belongs to the Old Testament, to the Second Book of Maccabees (7:28). Let us recall in passing that the two books of Maccabees, included in the Jerusalem Bible, are written in Greek and are not part of the Hebrew Bible. The text in question, where the author exhorts his son to take proper measure of the world, is translated as follows:

"I adjure you, my child, look at the heavens and the earth and see all that is in them, and know that God made them from nothing, and that the human race is made in the same way."

In reality, the Greek says: "I adjure you, son, concerning the things seen in heaven and on earth, and all the things seen in them, consider that God (o theos) did not make them from beings (ouk ex t\^{o}n ont\^{o}n epoiesen) and that the human race was born (ginetai) in this way." (see the Greek in \cite{The}).

The same negative turn of phrase is found in the second text invoked to justify the idea of "creation ex nihilo." This is a passage by Paul of Tarsus from the Epistle to the Romans (4:17), which speaks of Abraham and his unwavering faith in God, that is, as the Jerusalem Bible incorrectly translates,

"He in whom he believed, the God who gives life to the dead and calls nothingness into existence."

Here again, the Greek does not exactly say "calls nothingness into existence," but rather: "calling (kalountos) those who are not like those who are (ta m\`{e} onta \^{o}s onta)." The translation "to call into existence" for the Greek verb kale\^{o} is a pure invention, kale\^{o} simply meaning "to call," hence "to call to oneself," "to summon," or "to invite," but certainly not "to call into being," even if, in the context, it refers to "non-beings" (the English Bible, moreover, translates kale\^{o} as call).

The commentary of specialists on the subject supports our thesis. What is not or not yet is identified by Jewish thought with chaos or formless matter, and this is still the case in Greek Judaism, as, for example, in the Book of Wisdom (Wis 11:17) or in Philo of Alexandria (see \cite{May}, 10-11). The doctrine of "creatio ex nihilo" is now known as a late dogma, which only appears in the 2nd century AD, in Theophilus and Irenaeus, in the context of Gnostic doctrine and the influence of Greek philosophy, when it comes to defending the idea of an absolute transcendence of God in the face of various attacks and heresies (see also \cite{Fan}, 9). The debate will end with the anti-Gnostic theology of the second half of the 2nd century and the beginning of the 3rd century, when "creatio ex nihilo" officially becomes a fundamental doctrine of Christian theology.

It is the latter that we find at work in Augustine of Hippo in many of his works.\footnote{Augustine's most significant analyses concerning cosmology are found in his commentaries on Genesis ({\it De Genesi contra Manichaeos; De Genesi ad Litteram imperfectus liber; De Genesi ad Litteram-Libri Duodecim}). But, various remarks and investigations are still found in the last three books of the {\it Confessiones}, as well as in the {\it De Civitate Dei}, notably in books 11 and 12.}, and notably from the {\it Soliloquia} (I, 1, 2)(see \cite{Pri}). Subsequently, the reflection of Thomas Aquinas will develop the question in relation to that of time and eternity. "Everything that is created is made from nothing," writes Thomas. But everything that is made from nothing is a being after having been a nothing, since it is not simultaneously being and non-being" (see \cite{Tho}, 95-96). And although Thomas Aquinas considered possible, both logically and actually, the creation by God of a world without beginning (see \cite{Cel}, 318), the idea that something could exist outside of God "as if there could be something which nevertheless was not made by him" would be, according to the philosopher, an "abominable error" (see \cite{Tho}, 235). What is sometimes called an "agnosticism" of Thomas Aquinas is therefore quite partial, even if it is true that Thomas admitted that "the fact that the world has not always been is held only by faith, and cannot be proven demonstratively" (see \cite{Tho}, 196). As we know, this latter thesis, inspired by Maimonides and defended by Thomas against Bonaventure, ultimately won the day -- the great 13th-century controversy over the eternity of the world ended with his victory, which was not a solitary one. Ghent and Ockham, joining Boethius and Peckham, would join him in defending the possibility of the creation of an eternal world (even if faith asserted otherwise) (see, on this entire quarrel, \cite{Mic}). In view of this history, however, nothing obliges scientific cosmology to position itself in relation to this idea of creation from nothing, still less to seek complicated loopholes to circumvent what is, in reality, only an invention of the Fathers of the Church. Cyrille Michon, moreover, suggests, via the beginning of the Gospel of John, another interpretation of the Hebrew "berechit" (At the head). "The {\it principium} could not be understood in the sense of {\it chronological} beginning, but in the sense of {\it ontological} principle, and this principle would be the Word of God: it is in this principle, in the Word, that everything was made." (see \cite{Mic}, 18). But as the author acknowledges, this ancient exegesis had not succeeded in eliminating the chronological sense in the 13th century, and it is doubtful whether it has any more power to do so today.

\subsection{The So-Called "Harmony" of the World}

"Two things fill the heart (Gem\"{u}th) with ever new and ever increasing admiration and veneration, the more reflection is attached to them and involved in them," wrote Kant at the end of the {\it Critique of Practical Reason}: "the starry sky above me and the moral law within me" (see \cite{Kan2}, 172). Kant, in his time, could probably only distinguish about 3000 stars at most.\footnote{The number of stars accessible to the naked eye in the northern hemisphere is, theoretically, of the order of 5000, but under optimal conditions. Moreover, the power of the eye varies among individuals. In any case, he is unable to distinguish a flux of less than 50 photons per second, which roughly corresponds to the flux received at the back of the eye by a magnitude 6 star. But -- from experience -- distinguishing stars of magnitude 7.5 is already difficult!} of which he perhaps only knew a few dozen. This is obviously only a tiny fraction of the approximately 200 billion stars in the Milky Way. And let's not even mention the $10^{23}$ stars that are supposed to populate the Universe.

Consequently, the discovery of other galaxies (whose existence Kant, incidentally, had anticipated) and the true immensity of our Universe could obviously only reinforce this feeling of admiration and veneration that the German philosopher expressed in such a touching and almost romantic manner. From Johannes Kepler to Michio Kaku, including Einstein, Millikan, Hoffmann, Davies, Dyson and many others, countless physicists have believed in a divine order or, at the very least, a general harmony of the cosmos. Trihn Xuan Tuan seems to express the common opinion of this movement by estimating that, faced with the precise regulation that our universe seems to suppose, "we must postulate the existence of a Creative Principle which is at the origin of this regulation." (see \cite{Tri1}, 81). Elsewhere, he would express his innermost thoughts more precisely: "I reject the hypothesis of chance, because, even apart from the nonsense and despair it entails, I cannot conceive that the harmony, symmetry, unity, and beauty we perceive in the world, from the delicate contours of a flower to the majestic architecture of galaxies, but also (...) in the laws of Nature, are the result of chance alone. If we accept the hypothesis of a single universe, our own, we must postulate the existence of a First Cause that immediately established the physical laws and initial conditions for the Universe to become aware of itself." (see \cite{Tri2}, 445-446).

However, this finalism, which here takes the form of the Strong Anthropic Principle, is an epistemological principle proposed by astrophysicist Brandon Carter in 1974 (see \cite{Car}), is a principle that comes in two main versions. The weak anthropic principle ensures that what we can expect to observe must be compatible with the conditions necessary for our presence as observers (otherwise we would not be here to observe it). The strong anthropic principle postulates that the fundamental parameters on which the Universe depends are set once and for all so that it allows the birth and development of observers within it at a certain stage of its development. is subject to much criticism: Spinoza once urged us to be wary of this kind of fantasy: "the extravagance of men has gone so far as to believe that God also delights in harmony. There is no shortage of Philosophers who have convinced themselves that the celestial movements compose a harmony" (see \cite{Spi}, 66). He concluded: "We see that all the notions by which the common people are accustomed to explaining Nature are only Modes of imagining and do not inform us about the nature of anything, but only about the way in which the imagination is constituted (...)" (see \cite{Spi}, 67). Also, order and beauty are less in nature than in our mind, satisfied if it can easily represent things and memorize them (order) or if it can relate the movements it receives from the nerves, rather to health than to illness (beauty). But other notions such as Good, Evil, Hot, Cold, Praise, Blame, Sin or Merit are susceptible to the same analyses.

Concerning Nature, it will be noted that biologists are often less blind than physicists. In a book in which he combats some preconceived ideas, Marc-Andr\'{e} S\'{e}losse, professor at the Museum of Natural History in Paris, shows that Nature, seen up close, is not so harmonious as that. In the living world, in any case, each individual tends above all to pursue his own goals without worrying about the rest, until death finally reaches him. Natural evolution, by its selection of the most adapted, undoubtedly gives the illusion of an immense adjustment, but this, in reality, is only approximate and full of exceptions. Thus, work on symbiosis, even if we gradually understand that it cannot be reduced to either parasitism or mutualism, does not provide very firm indications or unequivocal results. Olivier Perru notes that, from the earliest research in the 1880s, authors showed that "symbiosis cannot be defined by mutual benefit. Within a symbiosis, certain unidirectional benefits can manifest themselves which tend to recall parasitism, even if there are obvious interrelations" (see \cite{Per1}, 27). It was not until the 1960s that serious cytological and genetic studies were carried out, and until the 1970s that a symbiotic theory of the cell and its evolution emerged (see \cite{Per2}). But the descent into the gene seems to show that nature does not function under the regime of altruism (see \cite{Daw}). As Selosse notes, following Dawkins, we imagine that insects and flowers have entered into a kind of mutual aid contract: transporting pollen in exchange for nectar. "Thus, a pedal system powders pollen from insects visiting young sage flowers, while later the older flowers scrape this pollen from the precise spot felt by the stamens." But this pattern admits of exceptions. The hawk-moth, a short-winged butterfly capable of hovering, absorbs nectar directly by means of a long pump that eliminates the need for it to collect pollen. The bumblebee, for its part, pierces the tube of the floral corolla with its mandibles and, likewise, collects nectar without taking or depositing pollen, allowing other insects to enter through the orifice thus created. On the flower side, the orchid does attract insects with its scent, but it is in vain for them to go pollinating it: it has no nectar. On the one hand, the insect's instinct leads it to feed without regard for the consequences for the plant (if necessary, it will shred it), and on the other, some plants have no regard for insects (which they unduly stimulate). Consequently, it is only too true that "nature is not made by or for anyone. It is simply, logically, populated by that which survives and reproduces, whatever the cost" (see \cite{Sel}, chap. 1). Nature, supposedly harmonious, is in fact mostly expensive, and the adaptation everyone raves about was achieved at the cost of a gigantic charnel house.

On a cosmological level, some believe the same is true, although, at that level, we cannot perceive the waste of waste. But for multiverse theorists, our universe, which managed to expand and produce roughly stable structures for a certain period of time, is only one of countless possible universes, some of which existed for only a few moments and others that remained desperately empty. The cosmic constants seem adjusted to us because we are precisely in the universe that allowed the development leading to man, but we do not see all the others -- the failed abortions (from our anthropomorphic point of view), but which could well exist in parallel, if we believe quantum cosmology.

\subsection{Conclusion}

Scientific cosmology, as it has developed, has gradually come to address metaphysical questions that were once the domain of philosophy or theology. The models capable of scientifically explaining the Universe as a whole are, admittedly, numerous, and some of their scenarios are contradictory. This is because calculations lead to different possible outcomes based on constants whose values can only be determined by observation -- observation that is still insufficient, but which is being refined and progressed over time and with technological developments (optical telescopes with increasingly large apertures, radio telescopes, space telescopes, etc.).

Each time, the harvest of new information allows us to obtain fundamental insights into the past of the Universe, or at least into its observable part. Although the cosmological horizon, the light horizon, and the Planck horizon constitute limits to our investigation, a detailed study of the 2.7 K background radiation may allow us to learn more and discover traces of the "Big Bang" or what preceded it.

In any case, scientific cosmology -- without completely excluding philosophy or even the mythologies of the past -- remains a science whose theories, duly mathematized, can only be taken seriously if supported by solid observations.

But for now -- and even in the eyes of believers who, in fact, know no more than anyone else -- the "meaning" of this entire construct (insofar as the question of meaning itself has meaning) continues to elude us, which is perhaps good news.

In his science fiction novel, {\it The Black Cloud}, Fred Hoyle humorously suggested that the answer to the ultimate metaphysical question -- the one humanity poses concerning the existence of an intelligence on a scale other than its own -- could be answered by the simple existence of an interstellar gas cloud. This might not be entirely surprising today, since, after all, the implementation of intelligence on structures other than carbon-based ones might well prove possible, at least in the eyes of proponents of a "functionalist" conception of the mind.

But, towards the end of the book, the intelligent cloud reveals that it itself is pondering the existence of intelligences greater than its own. To its knowledge, there are none in this galaxy or in any other galaxies it has encountered. But he admits that such intelligence could play a vital role in explaining what is: otherwise, how can we understand the behavior of matter and the nature of physical laws (why these, rather than others?).

According to the cloud itself, these problems are difficult to solve, but there is proof that their solution is possible. And to recount three disturbing events: two thousand million years ago, one of the cloud's fellow creatures claimed to have discovered an answer to the great metaphysical question. But the transmission of the message he then sent to the others was abruptly interrupted, and no one was able to restore it. Subsequently, no physical trace of the individual who sent it was found. Still according to the cloud, the same event would have occurred again 400 million years ago: a triumphant message, then... nothing, and no trace either of the one who sent it. But now the cloud has decided to leave the solar environment where it recharges near Earth because, two light-years away, a third event, of the same nature as the previous ones, has just occurred. The cloud therefore wants to move to the area, particularly with the aim of resolving a long-standing controversy among its peers: "It has been suggested, implausibly in my opinion, that these singular phenomena result from an abnormal neurological condition followed by suicide. It is not uncommon for suicide to take the form of a vast nuclear explosion causing the total disintegration of the individual. If this has occurred, then the failure to discover material traces of the individuals involved in these strange cases might be explained. In the present case, it should be possible for me to put this theory to the decisive test, for the incident, whatever it may have been, occurred so close that I will be able to reach the scene in only two or three hundred years. This is such a short time that the debris of the explosion, if there was an explosion, should not have entirely dispersed by then" (see \cite{Hoy}, chap. 12). By transposing the great metaphysical question to the level of "clouds," Hoyle thus humorously suggests that the solution we believe we have found could well correspond to a mental disorder, and that such a disorder inevitably leads to suicide by nuclear explosion. Clearly, the English astrophysicist wants us to understand that this is also what awaits humanity. Men who foolishly believe they have resolved the great metaphysical question are also on the brink of nuclear war. Suffice to say that, in this context, it would be better to leave the great metaphysical question open and avoid tearing each other apart over God, Allah, or any other imaginary entity, the imbecilic beliefs that go with it poisoning human life.


\begin{thebibliography}{}\addcontentsline{toc}{chapter}{Bibliography}

\bibitem[Adams-Laughlin 20]{Ada} Adams, F., Laughlin, G., {\it The Five Ages of the Universe: Inside the Physics of Eternity}, Free Press, New York, 2000.

\bibitem[An et al. 22]{An} An, D., Meissner, K. A.,  Nurowski, P. and Penrose, R., "Apparent evidence for Hawking points in the CMB",  {\it ArXiv}:1808.01740v5 [astro-ph.CO] 9 August 2022.

\bibitem[Bachelard 72]{Bac} Bachelard, G., "Univers et r\'{e}alit\'{e}", {\it L'engagement rationaliste}, 103-108, P.U.F., Paris, 1972.

\bibitem[Ba\~{n}ados et al. 18]{Ban} Ba\~{n}ados, E., Venemans, B., Mazzucchelli, C. et al., "An 800-million-solar-mass black hole in a significantly neutral Universe at a redshift of 7.5", {\it Nature} 553, 473-476 (2018).

\bibitem[Barrau et al. 10]{Bar1} Barrau, A. (et al.), {\it Multivers. Les mondes multiples de l'astrophysique, de la philosophie et de l'imaginaire}, Paris, La Ville b\^{u}le, 2010.

\bibitem[Barrau-Parrochia 10]{Bar2} Barrau, A., Parrochia, D. (ed.), {\it Forme et origine de l'Univers, regards philosophiques sur la cosmologie}, Dunod, Paris, 2010.

\bibitem[Bekenstein 73]{Bek} Bekenstein, J. D., "Black Holes and Entropy", {\it Phys. Rev.}, D 7, 2333-2346, 1973.

\bibitem[Caplan 20]{Cap} Caplan, M. E., "Black Dwarf Supernova in the Far Future", {\it arXiv}: 2008.02296 [astro-ph.HE], 2020.
 	
\bibitem[Carter 74]{Car} Carter, B.,  "Large number coincidences and the anthropic principle in cosmology", in M. Longair (ed.), {\it Confrontation of cosmological theories with observational data}, 291-298. D. Reidel, Dordrecht, 1974.

\bibitem[Celier 20]{Cel} Celier, G., {\it Saint Thomas d'Aquin et la possibilit\'{e} d'un monde cr\'{e}\'{e} sans commencement}, Via Romana, Versailles, 2020.

\bibitem[Dawkins 03]{Daw} Dawkins, R., {\it Le G\`{e}ne \'{e}go\"{i}ste}, Odile Jacob, Paris, 2003.

\bibitem[Deutsch 03]{Deu} Deutsch, D., {\it L'{e}toffe de la r\'{e}alit\'{e}} (1997), Paris, Cassini, 2003. 

\bibitem[d'Inverno-Vickers 22]{Din} d'Inverno, R., Vickers, J. {\it  Introducing Einstein's relativity : a deeper }, Oxford, Oxford University Press, 1992, rep. 2022

\bibitem[Eddington 34]{Edd} Eddington, A., {\it L'univers en expansion}, tr. fr. Hermann et Cie editeurs, Paris, 1934.

\bibitem[Dyson 79]{Dys}, Dyson, F., "Time without end: Physics and biology in an open universe", {\it Reviews of Modern Physics}, Vol. 51, No. 3, 447-460,1979.

\bibitem[Egan et al. 10]{Ega} Egan, C.A., Lineweaver, H., "A larger Estimate of the Entropy of the Universe", {\it Astrophysical Journal}, 710, 1825-1834, 2010. 

\bibitem[Fantino 96]{Fan} "L'origine de la doctrine de la cr\'{e}ation ex nihilo. \`{A} propos de l'ouvrage de G. May", {\it Revue des sciences philosophiques et theologiques}, 80, 589-602, 1996.

\bibitem[Friedman-Lema\`{i}tre 97]{Fri} Friedman, A., Lema\^{i}tre, G., {\it Essais de cosmologie}, pr\'{e}c\'{e}d\'{e} par "L'invention du Big Bang" par Jean-Pierre Luminet", Seuil, Paris, 1997.

\bibitem[Gasperini-Veneziano 03]{Gas1} Gasperini, M. ; Veneziano, G., "The pre-big bang scenario in string cosmology", {\it Physics Reports}, Volume 373, Issue 1-2, p. 1-212.

\bibitem[Gasperini-Veneziano 03]{Gas2} Gasperini, M. ; Veneziano, G., "Non-singular pre-big bang scenarios from all-order $\alpha'$ corrections", {\it arXiv}: 2305.00222v3 [hep-th] 19 Jul 2023.

\bibitem[Gupta 23]{Gup} Gupta, R. P., "JWST early Universe observations and $\Lambda$CDM cosmology", {\it Monthly Notices of the Royal Academic Society (MNRAS)} 524, 3385-3395, 2023.

\bibitem[Gurzadyan-Penrose 10]{Gur1} Gurzadyan, V. G., Penrose R., "Concentric circles in WMAP data may provide evidence of violent pre-Big-Bang", {\it ArXiv}, Novembre 2010.

\bibitem[Gurzadyan-Penrose 15]{Gur2} Gurzadyan, V. G., Penrose R., "CCC and the Fermi paradox",  {\it arXiv} :1512.00554v2 [astro-ph.CO], 14 D\'{e}c 2015.

\bibitem[Hartle-Hawking 83]{Hart}  Hartle J., et Hawking, S., "Wave function of the Universe", {\it Physical Review D}, vol. 28, No 12, 2960, 1983.

\bibitem[Hawking et al. 00]{Haw1} Hawking, S., Hertog, T. et Reall, H., "Brane new world", {\it  Physical Review D}, vol. 62, No 4, 043501, 2000, arXiv hep-th/0003052.

\bibitem[Hawking 74]{Haw8} Hawking, S., "Black holes aren't black", Wellesley, Gravity Research Foundation, 1974. 

\bibitem[Hawking 89]{Haw2} Hawking, S., {\it Une br\`{e}ve histoire du temps}, tr. fr. Flammarion, Paris, 1989, rep. 2020 et 2022.

\bibitem[Hawking 92]{Haw3} Hawking, S., {\it Commencement du temps et fin de la physique?}, tr. fr. Flammarion, Paris, 1992.

\bibitem[Hawking 94]{Haw4} Hawking, S., {\it Trous noirs et b\'{e}b\'{e}s univers}, tr. fr. Flammarion, Paris, 1994.

\bibitem[Hawking 01]{Haw7} Hawking, S., {\it L'univers dans une coquille de noix}, tr. fr. Odile Jacob, Paris 2001.

\bibitem[Hawking-Penrose 03]{Haw5} Hawking, S., Penrose, R., {\it La nature de l'espace et du temps}, tr. fr. Gallimard, Paris, 1997, folio, 2003.

\bibitem[Hawking-Mlodinow 14]{Haw6} Hawking, S., Mlodinow, L.., {\it Y a-t-il un grand architecte dans l'Univers?}, tr. fr. Odile Jacob, Paris, 2011, Poche, 2014.

\bibitem[Hoyle 57]{Hoy} Hoyle, F., {\it The Black Cloud}, Heinemann, Londres, 1957.

\bibitem[Kant 68]{Kan1} Kant, E., {\it Critique de la Raison pure}, tr. fr. P.U.F, Paris, 1968.

\bibitem[Kant 16]{Kan2} Kant, E., {\it Critique de la Raison pratique}, tr. fr. P.U.F., Paris, 2016.

\bibitem[Koyr\'{e} 62]{Koy2} Koyr\'{e}, A., {\it Du monde clos \`{a} l'univers infini}, Galllimard, Paris, 1962.

\bibitem[Koyr\'{e} 71]{Koy1} Koyr\'{e}, A., "Du monde de l'\`{a}-peu-pr\`{e}s \`{a} l'univers de la pr\'{e}cision", {\it Etudes d'histoire de la pens\'{e}e philosophique}, 341-362, Gallimard, Paris, 1971.

\bibitem[Labb\'{e} et al. 23]{Lab} Labb\'{e}, I., van Dokkum, P., Nelson, E. et al., "A population of red candidate massive galaxies $\sim$ 600 Myr after the Big Bang", {\it Nature} 616, 266-269 (2023).

\bibitem[Lachi\`{e}ze-Rey 96]{Lac} Lachi\`{e}ze-Rey, M., {\it Initiation \`{a} la cosmologie}, Masson, 2$^e$ ed., Paris, 1996.

\bibitem[Larson et al. 23]{Lar} Larson R. L. (et al.), "A CEERS Discovery of an Accreting Supermassive Black Hole 570 Myr after the Big Bang: Identifying a Progenitor of Massive z > 6 Quasars", {\it The Astrophysical Journal Letters}, Volume 953, Issue 2, id.L29, 1-26, 2023.

\bibitem[Linde 90]{Lin} Linde, A. D., {\it Inflation and Quantum Cosmology}, Academic Press Inc., San Diego, 1990.

\bibitem[Lopez et al. 22]{Lop1} Lopez, A. M., Clowes , R. G., et Willinger, G. M., "A Giant Arc in the Sky", {\it arXiv}:2201.06875v2 [astro-ph.CO], 2 August 2022.

\bibitem[Lopez et al. 24]{Lop2} Lopez, A. M., Clowes , R. G., et Willinger, G. M., "A Big Ring on the Sky", {\it arXiv}: 2402.07591 [astro-ph.CO], 12 February 2024.

\bibitem[Luminet et al. 03]{Lum}  Luminet, J.-P., Weeks, Riazuelo, A., Lehoucq, R., et Uzan, J.-P., "Dodecahedral space topology as an explanation for weak wide-angle temperature correlations in the cosmic microwave background", {\it Nature}, vol. 425, 593-595, 9 octobre 2003.

\bibitem[Marion 82]{Mar} Marion, J.-L., {\it Dieu sans l'\^{e}tre}, Communio/Fayard, Paris, 1982.

\bibitem[May 94]{May} G. May, {\it Creation ex nihilo. The Doctrine of "Creation out of Nothing" in Early Christian Thought}, T \& T Clarck Ltd, Edimbourg, 1994.

\bibitem[Merleau-Ponty 65]{Mer} Merleau-Ponty, J., {\it Cosmologie du XX$^e$ si\`{e}cle, \'{e}tude \'{e}pist\'{e}mologique et historique des th\'{e}ories de la cosmologie contemporaine}, Paris, Gallimard, 1965.

\bibitem[Michon 04]{Mic} Michon, C., {\it Thomas d'Aquin et la controverse sur l'\'{e}ternit\'{e} du monde}, Garnier-Flammarion, Paris, 2004.

\bibitem[Mugler 53]{Mug} Mugler, C., {\it Deux th\`{e}mes de la cosmologie grecque - Devenir cyclique et pluralit\'{e} des mondes}, Klincksieck, Paris, 1953.

\bibitem[Nesle-Aland 69]{Nes} Nestle, E., Aland, K., {\it Novum Testamentum Graece et Latine}, United Bible Society, London, 1963/1969.

\bibitem[Neves 17]{Nev} Neves, J. C. S., "Bouncing cosmology inspired by regular black holes", {\it Gen. Relativ. Gravit.} 49, 124, 2017.

\bibitem[Novello et al. 08]{Nov} Novello,M., Perez Bergliaffa, S. E., "Bouncing cosmologies", {\it Phys. Rept.} 463, 127, 2008.

\bibitem[Parrochia 19]{Par1} Parrochia, D., "Some remarks on history and pre-history of Feynman path integral", {\it arXiv}:1907.11168v1 [physics.hist-ph], 20 juin 2019.

\bibitem[Parrochia 21]{Par2} Parrochia, D., {\it Antidote ; contre le climato-dogmatisme et les discours apocalyptiques}, Le Corridor Bleu, 2021.

\bibitem [Penrose 06]{Pen1} Penrose, R., "Before the Big Bang: an outrageous new perspective and its implications for particle physics", {\it Proceedings of EPAC 2006}, 2759-2762, Edinburgh, Scotland, 2006.

\bibitem[Penrose 13]{Pen2} Penrose, R., {\it Les Cycles du temps : une nouvelle vision de l'Univers}, tr. fr. Odile Jacob, Paris, 2013. 

\bibitem[Perru 03]{Per2} Perru, O., {\it De la soci\'{e}t\'{e} \`{a} la symbiose, une histoire des d\'{e}couvertes sur les associations chez les \^{e}tres vivants, volume 1 : 1860-1930}, Vrin/IIEE Lyon, Paris, 2003.

\bibitem[Perru 06]{Per1} Perru, O., "Aux origines des recherches sur la symbiose vers 1868-1883", {\it Revue d'histoire des sciences}, Tome 59, 5-27, 1, 2006.

\bibitem[Peter-Uzan 05]{Pet} Peter, P., Uzan, J.-Ph., {\it Cosmologie primordiale}, Belin, Paris, 2005.

\bibitem[Pius XII 51]{Pie} Pius XII, {\it Discours \`{a} l'Acad\'{e}mie Pontificale des Sciences, le 22 novembre 1951}, La Documentation catholique, 33e ann\'{e}e, No 48, 16 d\'{e}cembre 1951.

\bibitem[Pricopi 18]{Pri} Pricopi, V. A., "Augustine of Hippo on creatio ex nihilo", AGATHOS, Volume 9, Issue 1, 16, 35-44, 2018.

\bibitem[Redford 11]{Red} Redford, {\it The Physics of God and the Quantum Gravity Theory of Everything}, Social Science Research Network, 2011.

\bibitem[Reeves 95]{Ree} Reeves, H., {\it La premi\`{e}re seconde - Derni\`{e}res nouvelles du cosmos, 2}, Seuil, Paris, 1995.

\bibitem[Robredo 06]{Rob} Robredo, J.-F., {\it Du cosmos au big bang - La r\'{e}volution philosophique}, P.U.F., Paris, 2006.

\bibitem[Scholem 56]{Sch} Scholem, G., "Sch\"{o}pfung aus nichts und Selbstverschr\"{a}nkung Gottes", {\it Eranos Jahrbuch} 25, 1956, p. 95-99.

\bibitem[Sechan 54]{Sec} Sechan, L., "Devenir cyclique et pluralit\'{e} des mondes chez les Grecs", {\it Les Etudes philosophiques}, Nouvelle S\'{e}rie, 9e Ann\'{e}e, No. 3, 360-365, Juillet/Septembre 1954.

\bibitem[Selosse 24]{Sel} Selosse, M.-A., {\it Nature et Pr\'{e}jug\'{e}s : convier l'humanit\'{e} dans l'histoire naturelle}, Acte Sud, Arles, 2024.

\bibitem[Silk 97]{Sil1} Silk, J., {\it Le Big Bang}, tr. fr. Odile Jacob, Paris, 1997.

\bibitem[Silk 03]{Sil2} Silk, J., {\it Une br\`{e}ve histoire de l'univers}, Odile Jacob, Paris, 2003.

\bibitem[Spinoza 65]{Spi} Spinoza, B., {\it \OE uvres III, Ethique}, tr. fr. Garnier-Flammarion, Paris, 1965.

\bibitem[Steiner et al. 08]{Ste} Janzer, H. S., Steiner, F., "Do we live in a small Universe?", {\it arXiv}:0708.1420 [astro-ph], 2008.

\bibitem[Stoffel 22]{Sto} Stoffel, J.-F., "Saint Thomas d'Aquin et la possibilit\'{e} d'un monde cr\'{e}\'{e} sans commencement - Gr\'{e}goire Celier" (recension), {\it Nouvelle Revue Th\'{e}ologique}, 144-1, 2022.

\bibitem[Susskind 08]{Sus} Susskind, L., {\it Le Paysage cosmique - notre univers en cacherait-il des millions d'autres}, tr. fr. Gallimard, folio essais, 2008.

\bibitem[Theotex-Lecture]{The} {\it Texte grec et traduction de la Bible des septantes}, Deuxi\`{e}me livre des Maccab\'{e}es, https://theotex.org/septuaginta/2maccabees/2maccabees$_{-}$ 7.html.

\bibitem[Thomas d'Aquin 20]{Tho} Thomas d'Aquin, {\it L'\'{e}ternit\'{e} du monde}, introduction et traduction par G. Celier, Vrin, Paris, 2020.

\bibitem[Tipler 94]{Tip} Tipler, F. J., {\it The Physics of Immortality: Modern Cosmology, God and the Resurrection of the Dead}, New York, Doubleday, 1994.

\bibitem[Tod 03]{Tod} Tod, K. P., "Isotropic cosmological singularities: Other matter models", {\it Class.Quant.Grav.} 20, 521-534, 2003.

\bibitem[Tolman 31]{Tol1} Tolman, R. C., "On the Theoretical Requirements for a Periodic Behaviour of the Universe", {\it Phys. Rev.} 38, 1758; Novembre 1931.

\bibitem[Tolman 34]{Tol2} Tolman, R. C., {\it Relativity, thermodynamics and cosmology}, Clarendon Press, Oxford, 1934.

\bibitem[Tresmontant 76]{Tre1} Tresmontant, C., {\it Sciences de l'univers et probl\`{e}mes m\'{e}taphysiques}, Seuil, Paris, 1976.

\bibitem[Tresmontant 07]{Tre2} Tresmontant, C., {\it L'histoire de l'univers et le sens de la cr\'{e}ation} (1985), Paris, Perrin, 2007.

\bibitem[Trinh Xuan Thuan 01]{Tri2} Trinh Xuan Thuan, {\it Le Chaos et l'Harmonie, la fabrication du r\'{e}el}, Fayard, Paris, 1998.

\bibitem[Trinh Xuan Thuan 01]{Tri1} Trinh Xuan Thuan, "La place de l'homme dans l'univers", in B. d'Espagnat (dir.), {\it Implications philosophiques de la science contemporaine}, tome 1, Le chaos, le temps, le principe anthropique, P.U.F., Paris, 2001.

\bibitem[Veneziano 04]{Ven} Veneziano, G., "L'univers avant le Big Bang", {\it Pour la Science}, N$^\circ$320, 40-47, Juin 2004.

\bibitem[Weinberg 88]{Wei} Weinberg, S., {\it Les trois premi\`{e}res minutes de l'univers}, Points-Seuil, Paris, 1988.

\bibitem[Wen et al. 24]{Wen} Wen, R. Y., Hergt, L. T., Afshordi, N. et Scott, D., "A cosmic glitch in gravity", {\it Journal of Cosmology and Astroparticle Physics}, 
 Volume 2024, Mars 2024, DOI 10.1088/1475-7516/2024/03/045.

\bibitem[Whittaker 46]{Whi} Whittaker, E., T., {\it Space and spirit - Theories of the Universe and the Arguments for the Existence of God}, Thomas Nelson, Londres, 1946. Rep. : Kessinger Publishing, Whitefish, MT, 2008.

\end{thebibliography}
\end{document}